\documentclass[10pt,twocolumn,showpacs,showkeys,preprintnumbers,amssymb,aps,superscriptaddress,prb]{revtex4-2}

\usepackage[utf8]{inputenc}
\usepackage{color}
\usepackage{amsmath}
\usepackage{graphicx}
\usepackage{hyperref}
\hypersetup{colorlinks=true,
 	linkcolor=blue,
 	citecolor=blue,
 	urlcolor=blue} 
\usepackage{dcolumn}
\usepackage{bm}
\usepackage{wrapfig}
\usepackage{subfigure}
\usepackage{ucs}
\usepackage{lineno}
\usepackage{epsfig}
\usepackage{psfrag}
\usepackage{epstopdf}
\usepackage{relsize}
\usepackage{amssymb}
\usepackage{pifont}
\usepackage{xcolor, pifont, wasysym}
\usepackage{tikz}
\usepackage{orcidlink}
\usepackage{pstricks}
\usepackage{comment}
\makeatletter
\@ifundefined{textcolor}{}
{
	\definecolor{BLACK}{gray}{0}
	\definecolor{WHITE}{gray}{1}
	\definecolor{RED}{rgb}{1,0,0}
	\definecolor{GREEN}{rgb}{0,1,0}
	\definecolor{BLUE}{rgb}{0,0,1}
	\definecolor{CYAN}{cmyk}{1,0,0,0}
	\definecolor{MAGENTA}{cmyk}{0,1,0,0}
	\definecolor{YELLOW}{cmyk}{0,0,1,0}
}

\begin{document}
	
\title{Cavity-QED enhancement of quantum entanglement and battery performance in double quantum dots}

\author{Hamid Arian Zad\orcidlink{0000-0002-1348-1777}}
\email{Corresponding author: arianzad.hamid@gmail.com}
\address{Department of Theoretical Physics and Astrophysics, Faculty of Science, P. J. \v{S}af{\'a}rik University, Park Angelinum 9, 040 01 Ko\v{s}ice, Slovak Republic}

\author{Usama Shoukat}
\affiliation{Department of Physics, Abbottabad University of Science and Technology, Havellian, 22500, Pakistan}


\author{Michal Ja{\v s}{\v c}ur\orcidlink{0000-0003-0826-1961}}
\address{Department of Theoretical Physics and Astrophysics, Faculty of Science, P. J. \v{S}af{\'a}rik University, Park Angelinum 9, 040 01 Ko\v{s}ice, Slovak Republic}

\author{Hazrat Ali~\!\!\orcidlink{0000-0003-1957-3629}}
\affiliation{Department of Physics, Abbottabad University of Science and Technology, Havellian, 22500, Pakistan}

\author{Saeed Haddadi\orcidlink{0000-0002-1596-0763}}
\address{School of Particles and Accelerators, Institute for Research in Fundamental Sciences (IPM), P.O. Box 19395-5531, Tehran, Iran}

\begin{abstract}
We theoretically investigate quantum correlations and energy storage in a single-electron silicon double quantum dot (eDQD) coupled to a single-mode microwave cavity. The charge and spin degrees of freedom are hybridized by Rashba spin-orbit coupling, while the cavity interacts with the eDQD through spin-photon and charge-photon couplings. Using the reduced thermal density matrix of the eDQD, we characterize spin-charge entanglement by the concurrence and quantum coherence by the \(l_1\) norm, and analyze their dependence on temperature, Rashba coupling, cavity frequency, and light--matter coupling strengths. At low temperature, the cavity strongly restructures the spin-charge correlations and produces distinct regimes of enhanced and suppressed entanglement. We identify a nonlinear crossover between an eDQD-dominated regime and a photon-dressed regime. This crossover is revealed independently by two signatures: a rapid change in the concurrence and the onset of finite photon occupation in the cavity. Its boundary exhibits a dominant \(G_{\rm c}\propto\sqrt{\Omega}\) dependence and is robust against cavity Hilbert-space truncation. In addition to the Rashba coupling studied in~[Ferreira \textit{et al.}, Phys. Rev.~A \textbf{107}, 052408 (2023)], we show that the cavity provides an additional tunable tool for simultaneously controlling quantum correlations and energy-storage properties. We further characterize the eDQD as a quantum battery under coherent charging and determine its stored energy and ergotropy. The ergotropy varies markedly across a cavity-dressing crossover resembling the one identified from the entanglement analysis, linking the modification of the dressed spin-charge-photon states to the extractable work. Optimization over the charge-photon coupling reveals parameter-dependent optimal charging regimes and an enhancement of the maximum ergotropy with increasing Rashba coupling. Our results establish cavity coupling and spin-orbit interaction as complementary control parameters for quantum correlations and extractable energy in semiconductor eDQD--cavity systems.
\end{abstract}

\maketitle
	
\section{Introduction}

Semiconductor double quantum dots (DQDs) provide a versatile solid-state platform for quantum information processing because of their electrical tunability, scalability, and compatibility with established semiconductor technology~\cite{Loss1998,Hanson2007,Kloeffel2013}. Tunnel coupling between two spatially localized electronic states gives rise to an artificial molecular system whose orbital and spin degrees of freedom can be exploited for quantum-state encoding and manipulation~\cite{Hayashi2003,Petta2005}. For a single electron confined in a DQD (eDQD), the localized states $|L\rangle$ and $|R\rangle$ define a charge pseudospin, while the electron spin constitutes a second two-level degree of freedom, resulting in a four-dimensional spin-charge Hilbert space~\cite{Ferreira2023}. The charge sector offers fast electrical control, while coupling between the orbital and spin degrees of freedom enables the generation and manipulation of nonclassical spin-charge states. Understanding the quantum correlations and superpositions supported by this coupled system is therefore important for assessing the eDQD as a controllable quantum device.

Among the quantum resources relevant to eDQDs, entanglement and quantum coherence provide complementary information about the spin-charge state. For the two-qubit system, entanglement can be quantified by the concurrence measure~\cite{Wootters1998}, whereas coherence can be characterized by the $l_1$-norm of coherence in a specified basis using the off-diagonal elements of the density matrix ~\cite{Baumgratz2014}. Although coherence and entanglement are distinct quantum resources, they are closely related, and coherence can be converted into entanglement under suitable operations~\cite{Streltsov2015,Streltsov2017}. Importantly, $l_1$-norm of coherence may remain finite in mixed states for which concurrence has already vanished, allowing residual quantum superpositions to be identified beyond the temperature range in which spin--charge entanglement is detectable. The simultaneous analysis of concurrence and coherence therefore provides a more complete characterization of the robustness of quantum features in eDQDs, particularly under thermal fluctuations and coupling to additional degrees of freedom.

Rashba spin-orbit coupling (SOC), originating from structural inversion asymmetry~\cite{Rashba1960}, couples the spin and orbital degrees of freedom and enables electrical manipulation of electron spins. In DQDs, this interaction hybridizes the charge and spin states and thereby provides a direct mechanism for controlling spin-charge coherence and entanglement~\cite{Nadj2013,Sen2023,Svastits2026}. The interplay of SOC and magnetic fields in DQDs has been investigated in several settings~\cite{Sen2023,Leitao2025}, while recent studies have shown that Rashba coupling can strongly modify thermal entanglement, coherence, and related quantum information quantities in single-electron DQDs~\cite{Ferreira2023,Oumennana2024}. Such systems therefore provide a compact solid-state platform in which quantum correlations can be tuned by experimentally accessible parameters.

Silicon is particularly attractive for realizing DQD-based quantum devices because of its compatibility with mature fabrication technology and the availability of isotopically enriched $^{28}$Si, which strongly suppresses hyperfine-induced spin dephasing~\cite{Kloeffel2013,Zwanenburg2013}. High-fidelity control has consequently been demonstrated in Si/SiGe quantum-dot processors, including single- and two-qubit fidelities exceeding $99\%$~\cite{Mills2022,Xue2022}. Coupling silicon DQDs to superconducting microwave resonators further extends this platform by introducing a coherent interface between electronic and photonic degrees of freedom~\cite{Burkard2020,Burkard2023,Yu2023}. Within the rotating-wave approximation, the interaction of a two-level system with a single quantized electromagnetic mode is described by the Jaynes--Cummings model~\cite{Jaynes1963,Larson2024}. In the strong-coupling regime, coherent light--matter interaction can give rise to hybridized states and characteristic signatures such as vacuum Rabi splitting~\cite{Raimond2001,Toida2013,Blais2021}. Experimentally, coherent coupling between a single electron spin in a silicon DQD and a microwave cavity has been demonstrated~\cite{Mi2018}, together with strong spin--photon coupling in silicon hole-spin systems~\cite{Yu2023} and photon-mediated interactions between spatially separated DQD charge qubits~\cite{vanWoerkom2023}. These developments establish cavity-coupled DQDs as a controllable spin--charge--photon architecture.

Beyond quantum information and computing, controllable few-level quantum systems provide a natural setting for investigating energy transfer and storage at the quantum scale. A quantum battery (QB) \cite{Quach2023,Campaioli2024,Camposeo2025,Kurman2026,Le2018} stores energy in the internal degrees of freedom of a quantum system, from which useful work can subsequently be extracted, with the maximum extractable work quantified by the ergotropy under cyclic unitary operations. Quantum correlations can effectively influence charging dynamics and extractable work done by the QB~\cite{PhysRevA.110.052404,HaddadiAQT2025}, motivating extensive studies of QBs within quantum thermodynamics.
More recently, fundamental limitations on the reliability of QBs have been identified, revealing an unavoidable trade-off between charging-power fluctuations and the work deposited during charging and discharging~\cite{Mohan2026}. Nevertheless, these fundamental limitations make it particularly interesting to examine specific physical architectures and controllable mechanisms for energy transfer and storage, as they provide specific platforms for exploring how quantum correlations, dissipation, and light-matter interactions affect battery performance.
Experimental progress has begun to establish these concepts in physical platforms, including nuclear-spin systems, where quantum correlations have been associated with an experimentally observed charging advantage~\cite{Joshi2022}. Previously, a quantum-battery scheme comprising an atom-cavity interacting system coupled to a structured reservoir was investigated in Ref.~\cite{HaddadiAQT2024}. The results showed that dissipative effects reduce the energy stored in the battery, whereas stronger battery-cavity coupling, improves the charging performance and enhances the extractable work. Cavity-coupled eDQDs are especially appealing in this context because their internal spin and charge degrees of freedom can serve as an energy-storage subsystem while the cavity provides a controllable channel for energy exchange. In the architecture considered here, the eDQD constitutes the QB and a single-mode microwave cavity mediates its interaction with an auxiliary two-level charger. An initially excited charger can transfer energy coherently to the cavity, whose photons subsequently couple to the eDQD through spin--photon and charge--photon interactions. 

The Rashba SOC provides an internal pathway for redistributing energy and quantum coherence within the battery. In addition to this cavity-mediated route, the eDQD can be charged directly by a local external field, allowing the two charging mechanisms to be compared. The battery performance can then be characterized not only by the energy accumulated in the eDQD but also by its ergotropy, which distinguishes total stored energy from the fraction that is accessible as useful work.
The performance of such a hybrid QB is intrinsically connected to the quantum state of the eDQD. Thermal fluctuations and environmental noise can suppress the coherence and correlations that accompany the charging process. In semiconductor QDs, relevant decoherence channels include hyperfine coupling to nuclear spins~\cite{Khaetskii2003}, electrical charge noise~\cite{Kuhlmann2013,Dial2013}, cavity-photon loss~\cite{Benito2017}, and phonon-mediated relaxation~\cite{Burkard2023}. At finite temperature, thermal population of excited states further modifies the reduced spin-charge density matrix. Since entanglement can disappear at a finite temperature while coherence may persist over a broader thermal range, analyzing both quantities together with stored energy and ergotropy provides a more complete picture of the robustness and energetic usefulness of the eDQD state.

Motivated by these considerations, we investigate a single-electron DQD with Rashba SOC coupled to a single-mode microwave cavity, addressing both its quantum correlations and its performance as a cavity-mediated QB. After tracing out the cavity degrees of freedom, we characterize the reduced eDQD state through the concurrence and the $l_1$-norm of coherence and determine their dependence on temperature, Rashba SOC, cavity frequency, and spin- and charge-photon coupling strengths. We then employ the eDQD as the battery subsystem and investigate coherent energy transfer under local and cavity-mediated charging protocols. The charging performance is quantified through the stored energy and ergotropy, allowing us to distinguish energy accumulation from unitarily extractable work. This framework enables us to identify how Rashba-induced spin-charge hybridization and cavity coupling can be used to control quantum correlations, energy transfer, and useful energy storage in a semiconductor eDQD.

The paper is organized as follows. Section~\ref{sec:model} introduces the model of a single-electron DQD with Rashba SOC coupled to a microwave cavity, together with the total Hamiltonian and the thermal density matrix. In Sec.~\ref{Sec:ConCoh}, we present the behavior of the concurrence and $l_1$-norm coherence as functions of temperature, Rashba SOC, and the relevant cavity parameters, and analyze the relation between these quantum resources. Section \ref{Sec:DQDs_QB} is devoted to discuss the performance of a QB based on the eDQD interacting with a cavity QED. Finally, Sec.~\ref{sec:conclusion} summarizes the main conclusions and provides an outlook for future work.

\section{The Cavity-Embedded eDQD Model}
\label{sec:model}

\begin{figure}[t]
	\centering
	\resizebox{0.48\textwidth}{!}{
		\includegraphics[trim = 0 0 10 0, clip]{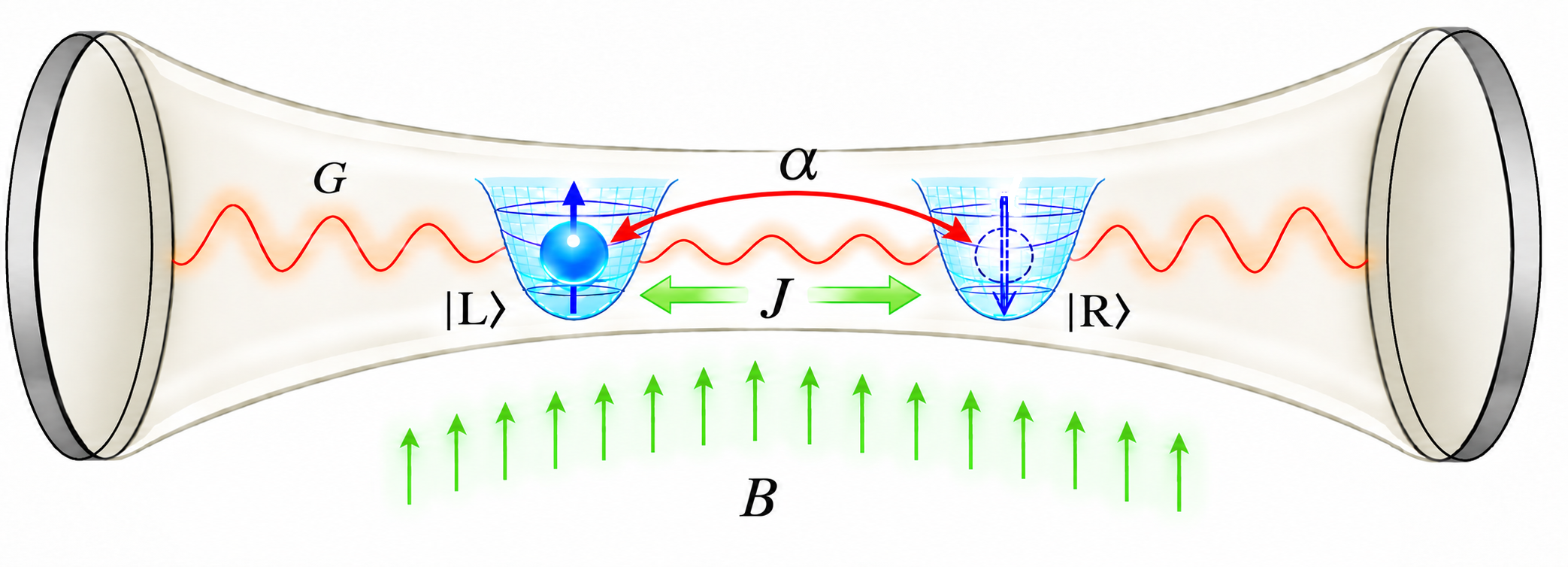}}
	\vspace{-0.5cm}
	\caption{
		Schematic representation of an eDQD embedded in a microwave cavity QED. The electron can tunnel between the left and right dots with amplitude $J$, experiencing Rashba SOC $\alpha$ that mixes spin and charge degrees of freedom. $B$ is the Zeeman splitting as a result of a homogeneous magnetic field along the $z$-axis. The cavity mode of frequency $\Omega$ couples to the electron spin via the Jaynes--Cummings interaction of strength $G$, enabling coherent spin-photon exchange.
	}
	\label{fig:DQDs_e_cQED}
\end{figure}

By inspecting Fig.~\ref{fig:DQDs_e_cQED}, we can generalize the Hamiltonian of the system with three distinct parts: eDQD with Rashba SOC, the quantized cavity field, and their mutual interaction. The eDQD consists of two tunnel-coupled quantum dots, where the electron can occupy either the left ($|L\rangle$) or right ($|R\rangle$) dot, forming a charge qubit, while its spin degree of freedom ($|\uparrow\rangle, |\downarrow\rangle$) constitutes a spin qubit. The cavity supports a single-mode electromagnetic field that mediates long-range interactions and provides a controllable environment for the eDQD system.

The hybrid eDQD-cavity system consists of three coupled physical degrees of freedom: the charge degree of freedom of the DQD, the electron spin, and a single quantized cavity mode. The total Hamiltonian can thus be written as 
\begin{equation}
	\hat{H}_{\mathrm{tot}} = \hat{H}_{\mathrm{eDQD}} + \hat{H}_{\mathrm{c}} + \hat{H}_{\mathrm{int}},
\end{equation}
where $\hat{H}_{\mathrm{eDQD}}$ describes the isolated eDQD, $\hat{H}_{\mathrm{c}}$ describes the quantized electromagnetic mode, and $\hat{H}_{\mathrm{int}}$ contains both spin-photon and charge-photon interactions. We consider a single electron confined in a DQD, with the two localized charge states denoted by $|L\rangle$ and $|R\rangle$, and the spin states by $|\uparrow\rangle$ and $|\downarrow\rangle$. The specific form of the Hamiltonian of the eDQD  reads

\begin{equation}\label{eq:H_eDQD}
	\hat{H}_{\mathrm{eDQD}} = J(\eta_x\otimes \mathbb{I}) + \frac{B}{2} (\mathbb{I}\otimes\sigma_z) - \alpha(\eta_y\otimes\sigma_x).
\end{equation}
Here, $\eta_i$ ($i=x,y,z$) are Pauli operators acting on the charge subspace $\{|L\rangle,|R\rangle\}$, whereas $\sigma_i$ are Pauli operators acting on the spin subspace $\{|\uparrow\rangle,|\downarrow\rangle\}$. The term $\mathbb{I}$ denotes the corresponding identity operator for either of charge or electron. The parameter $J$ is the coherent interdot tunneling amplitude and therefore controls the hybridization between the localized charge states. The parameter $\alpha$ characterizes the Rashba SOC, which couples charge motion to the electron spin through the term $\eta_y\otimes\sigma_x$. The parameter  $B=g\mu_\text{B}h_z$  determines the effective spin splitting associated with the $\sigma_z$ degree of freedom, where $\mu_{\mathrm{B}}$ is the Bohr magneton, $g$ is the gyromagnetic factor, and $h_z$ is the applied magnetic field along the $z$ direction.

The cavity is modeled as a single quantized harmonic-oscillator mode with 
\begin{equation}
	\hat{H}_{\mathrm{c}} = \hbar\Omega\,\hat{a}^{\dagger}\hat{a},
\end{equation}
in which $\hat{a}^{\dagger}$ and $\hat{a}$ are the photon creation and annihilation operators satisfying
$[\hat{a},\hat{a}^{\dagger}] = 1$, and $\Omega$ is the cavity frequency. The zero-point energy $\hbar\Omega/2$ is omitted since it only produces an overall energy shift and has no effect on the dynamics or on the extracted work.

The DQD can couple to the cavity through both its spin and charge degrees of freedom. Accordingly, the interaction Hamiltonian $\hat{H}_{\mathrm{int}}$ is written as
\begin{equation}
	\hat{H}_{\mathrm{int}} = G(\hat{a}^{\dagger}\sigma_- + \hat{a}\sigma_+) + g_\text{c}(\hat{a}+\hat{a}^{\dagger})(\eta_z\otimes \mathbb{I}),
\end{equation}
where the parameter $G$ denotes the spin-cavity (spin-photon) coupling strength, whereas $g_\text{c}$ is the charge-cavity coupling strength. 
The first interaction $G(\hat{a}^{\dagger}\sigma_- + \hat{a}\sigma_+)$ has the Jaynes-Cummings form. As a matter of fact, it couples the spin transition to the cavity field and enables coherent exchange between spin and photon excitations. The terms $\hat{a}^{\dagger}\sigma_-$ and $\hat{a}\sigma_+$ change the photon number and spin state simultaneously. In the rotating-wave approximation, this interaction conserves the corresponding excitation number within the spin-photon sector.
The second interaction term $g_\text{c}(\hat a+\hat a^\dagger)(\eta_z\otimes \mathbb{I})$ describes the electric-dipole coupling between the cavity field and the eDQD charge degree of freedom. In contrast to the Jaynes-Cummings interaction, this term contains both photon creation and annihilation processes and therefore does not conserve the total excitation number. Nevertheless, as shown below, it preserves a generalized $\mathbb{Z}_2$ parity symmetry when combined with the charge and spin transformations.

The coexistence of $G$ and $g_\text{c}$ is particularly important for the present hybrid QB architecture. The coupling $G$ provides a direct spin--photon channel, while $g_\text{c}$ provides a direct charge--photon channel. Because charge and spin are intrinsically linked by the Rashba interaction $\alpha$, the cavity can mediate a nontrivial interplay among charge, spin, and photonic degrees of freedom. This interplay can modify the quantum coherence, entanglement, and ultimately the amount of extractable work stored in the QB based on the eDQD.

The complete Hamiltonian can therefore be written explicitly as
\begin{align}\label{Eq:total_H}
	\hat{H}_{\mathrm{tot}}
	={}&
	J\left(\eta_x\otimes \mathbb{I}\right)
	+ \frac{B}{2}
	\left(\mathbb{I}\otimes\sigma_z\right)
	-\alpha\left(\eta_y\otimes\sigma_x\right)	+\hbar\Omega\,\hat{a}^{\dagger}\hat{a}\nonumber\\
	&+G\left(\hat{a}^{\dagger}\sigma_-+\hat{a}\sigma_+\right)
+g_\text{c}\left(\hat{a}+\hat{a}^{\dagger}\right)
	\left(\eta_z\otimes \mathbb{I}\right),
\end{align}
with dimensions $\dim(\mathcal{H}) = 2\times2\times N,$ where $N$ denotes the number of retained cavity Fock states. Since the model cannot be solved analytically, the Hamiltonian is then solved numerically using exact diagonalization.

\section{Quantum coherence and entanglement in eDQD}\label{Sec:ConCoh}

At finite temperature, the eDQD-cavity composite system can be described by the canonical Gibbs state
\begin{equation}
	\rho_{\mathrm{tot}}=\frac{e^{-\beta H_{\mathrm{tot}}}}{Z},
	\label{eq:gibbs}
\end{equation}
where $\beta=1/k_{\mathrm B}T$ is the inverse temperature, $k_{\mathrm B}$ is the Boltzmann constant, and $Z=\mathrm{Tr}\left[e^{-\beta H_{\mathrm{tot}}}\right]$ is the partition function. Since our interest is focused on the quantum correlations between the charge and spin degrees of freedom of the eDQD, the cavity degrees of freedom are traced out, yielding the reduced density matrix $\rho_{\mathrm{eDQD}}=\operatorname{Tr}_{\mathrm{c}}\left(\rho_{\mathrm{tot}}\right)$.
Here, the quantum properties of the charge-spin state are characterized by two complementary quantities including the $\ell_1$-norm of coherence and the concurrence. The former quantifies the magnitude of the off-diagonal elements of $\rho_{\mathrm{eDQD}}$ in the chosen charge-spin basis (\ref{Eq:eDQD_basis}) and is defined as~\cite{Baumgratz2014}
\begin{equation}
	C_{\ell_1}(\rho_{\mathrm{eDQD}})=\sum_{i\neq j}\left|(\rho_{\mathrm{eDQD}})_{ij}\right|,
	\label{eq:l1_coherence}
\end{equation}
whereas the concurrence quantifies the bipartite entanglement between the charge and spin degrees of freedom. For the two-qubit reduced eDQD state, the concurrence is given by~\cite{Wootters1998}
\begin{equation}
	C(\rho_{\mathrm{eDQD}})=\max\left(0,\lambda_1-\lambda_2-\lambda_3-\lambda_4\right),
	\label{eq:wootters_concurrence}
\end{equation}
where $\lambda_i$, arranged in descending order, are the eigenvalues of
\begin{equation}
	R=\sqrt{\sqrt{\rho_{\mathrm{eDQD}}}\,\widetilde{\rho}_{\mathrm{eDQD}}\,\sqrt{\rho_{\mathrm{eDQD}}}},
	\label{eq:R_matrix}
\end{equation}
with $\widetilde{\rho}_{\mathrm{eDQD}}=(\sigma_y\otimes\sigma_y)\rho_{\mathrm{eDQD}}^{*}(\sigma_y\otimes\sigma_y)$.
Here, $\rho_{\mathrm{eDQD}}^{*}$ denotes complex conjugation in the computational charge-spin basis. While $C_{\ell_1}$ characterizes coherence with respect to this fixed basis, the concurrence measures genuine charge-spin entanglement and ranges from $C=0$ for separable states to $C=1$ for maximally entangled two-qubit states. Therefore, their combined analysis allows us to distinguish the overall coherence of the reduced eDQD state from the specifically nonseparable quantum correlations between its charge and spin sectors.
 
The total Hamiltonian $\hat{H}_{\mathrm{tot}}$ governs the dynamics of the eDQD, the cavity mode, and their mutual interactions, as introduced in Sec.~\ref{sec:model}. The thermal state in Eq.~(\ref{eq:gibbs}) assumes that the system is in thermal equilibrium with its environment at temperature $T$. This description is appropriate for studying the equilibrium properties of coherence and entanglement.
To render the problem numerically tractable, we truncate the cavity Hilbert space to a maximum photon number $N$. The photon-number in the Fock basis is defined by the eigenvalue equation $\hat{a}^{\dagger}\hat{a}|n\rangle = n|n\rangle$ with $n = 0,1,\ldots,N$ and the orthonormality condition $\langle n|m\rangle = \delta_{nm}$.
The truncated cavity Hilbert space is then; 	$\mathrm{span}\left\{|0\rangle, |1\rangle, \ldots, |N\rangle\right\}$.
The truncation parameter $N$ must be chosen sufficiently large to ensure convergence of all physical observables of interest. Throughout this work, we increase $N$ until the relative variation in the computed quantities, such as quantum coherence and correlations, is negligible.

The eDQD subsystem can be described by the reduced density matrix obtained by tracing out the cavity degrees of freedom $\rho_{\mathrm{eDQD}} = \mathrm{Tr}_{\mathrm{c}}\left[\rho_{\mathrm{tot}}\right]$, where $\mathrm{Tr}_{\mathrm{c}}$ denotes the partial trace over the cavity subspace. In the Fock basis, this partial trace takes the explicit form
\begin{equation}
	\rho_{\mathrm{eDQD}} = \sum_{n=0}^{N} \langle n| \rho_{\mathrm{tot}} |n\rangle.
	\label{eq:partial_trace}
\end{equation}
To evaluate Eq.~(\ref{eq:partial_trace}), we employ the spectral decomposition of the total Hamiltonian. Let $\{|\Psi_{\lambda}\rangle\}$ and $\{E_{\lambda}\}$ denote the eigenstates and eigenenergies of $\hat{H}_{\mathrm{tot}}$ defined in Eq. (\ref{Eq:total_H}), satisfying the time-independent Schr\"odinger equation $\hat{H}_{\mathrm{tot}}|\Psi_{\lambda}\rangle = E_{\lambda}|\Psi_{\lambda}\rangle$, 
with the orthonormality condition $\langle \Psi_{\lambda}|\Psi_{\mu}\rangle = \delta_{\lambda\mu}$. The thermal density matrix in Eq.~(\ref{eq:gibbs}) can then be expressed spectrally as
\begin{equation}
	\rho_{\mathrm{tot}}(T) = \frac{1}{Z} \sum_{\lambda} e^{-\beta E_{\lambda}} |\Psi_{\lambda}\rangle \langle \Psi_{\lambda}|,
	\label{eq:spectral}
\end{equation}
where the partition function is given by $Z = \sum_{\lambda} e^{-\beta E_{\lambda}}$.
Substituting Eq.~(\ref{eq:spectral}) into Eq.~(\ref{eq:partial_trace}), the thermal reduced density matrix of the eDQD becomes
\begin{equation}
	\rho_{\mathrm{eDQD}}(T) = \frac{1}{Z} \sum_{\lambda} e^{-\beta E_{\lambda}} \, \mathrm{Tr}_{\mathrm{c}}\left[|\Psi_{\lambda}\rangle \langle \Psi_{\lambda}|\right].
	\label{eq:Reduced_rho_eDQD}
\end{equation}
It is important to emphasize that the index $n$ in Eq.~(\ref{eq:partial_trace}) labels the Fock basis states used in the partial trace, whereas $\lambda$ labels the eigenstates of the full interacting Hamiltonian. The eigenstates $|\Psi_{\lambda}\rangle$ are generally dressed eDQD-cavity states, containing contributions from multiple photon-number sectors. Consequently, the partial trace in Eq.~(\ref{eq:Reduced_rho_eDQD}) couples different photon sectors through the entanglement between the eDQD and the cavity mode.
In the localized spin-charge basis
\begin{equation}\label{Eq:eDQD_basis}
	\mathcal{B} = \left\{|\text{L}\uparrow\rangle, |\text{L}\downarrow\rangle, |\text{R}\uparrow\rangle, |\text{R}\downarrow\rangle\right\},
\end{equation}
the  thermal reduced density matrix $\rho_{\mathrm{eDQD}}(T)$ is a $4\times 4$ Hermitian, positive-semidefinite matrix,
with normalization condition $\sum_{i=1}^{4} \rho_{ii} = 1$.
This reduced density matrix provides the foundation for evaluating all quantum resource measures considered in this work, including concurrence, $l_1$-norm coherence, and ergotropy.

A non-zero concurrence in the eDQD system indicates that the spin and charge degrees of freedom are entangled, meaning that measurements on one subsystem can reveal information about the other. This entanglement is a fundamental resource for quantum communication and quantum computing.
Such a kind of entanglement is a manifestation of the hybrid spin-charge nature of the system and is generated by the Rashba SOC and cavity interactions.
The relationship between coherence and entanglement in this system is of particular interest. While coherence is a single-party property (it depends only on the reduced density matrix of the eDQD), entanglement is a multi-party property that requires correlations between subsystems. However, as demonstrated in the literature~\cite{Streltsov2015}, coherence in a system can be converted into entanglement through incoherent operations, establishing a deep connection between these two resources. Our analysis of both measures will reveal how these complementary aspects of quantum correlations behave under different parameter regimes and how they can be optimized for quantum information applications.

In this section, we present our numerical results for the $l_1$-norm coherence and concurrence of the cavity-embedded eDQD system. We examine how these quantum resources depend on the Jaynes--Cummings spin-photon coupling strength \(G/J\), the Rashba SOC \(\alpha/J\), the cavity frequency \(\Omega/J\), and the temperature \(k_\text{B}T/J\). Throughout, we work in natural units with \(\hbar = 1\), and we take the interdot tunneling amplitude $J$ as the energy unit.

\begin{figure*}[t]
	\centering
	\resizebox{0.327\textwidth}{!}{
		\includegraphics[trim = 0 0 10 0, clip]{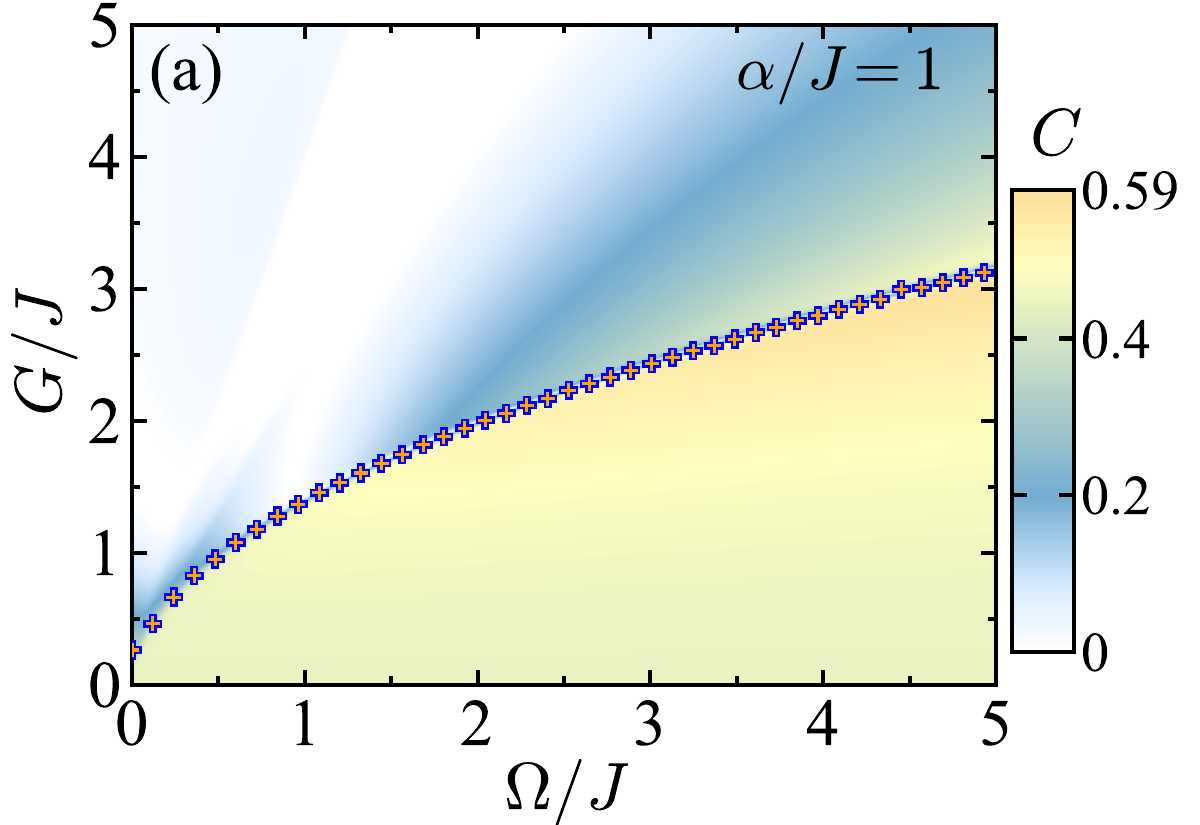}}
	\resizebox{0.327\textwidth}{!}{
		\includegraphics[trim = 0 0 10 0, clip]{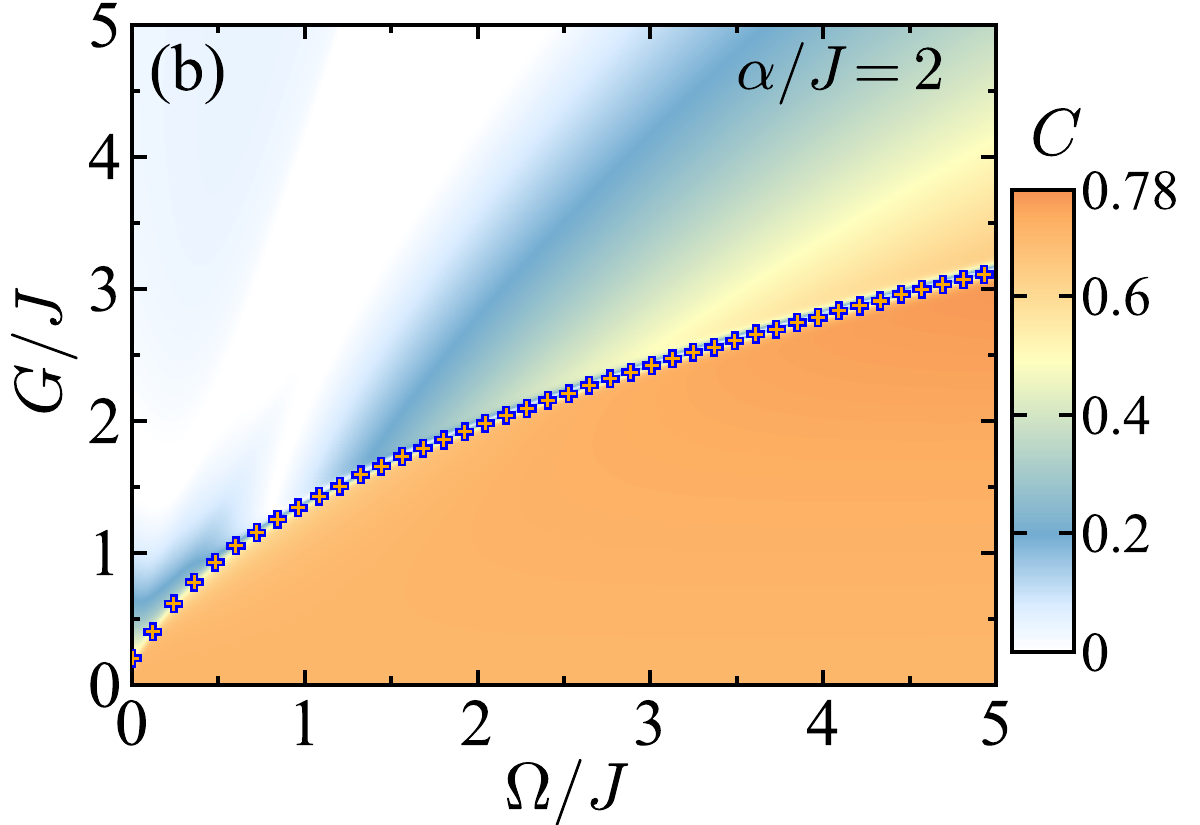}}
	\resizebox{0.327\textwidth}{!}{
		\includegraphics[trim = 0 0 10 0, clip]{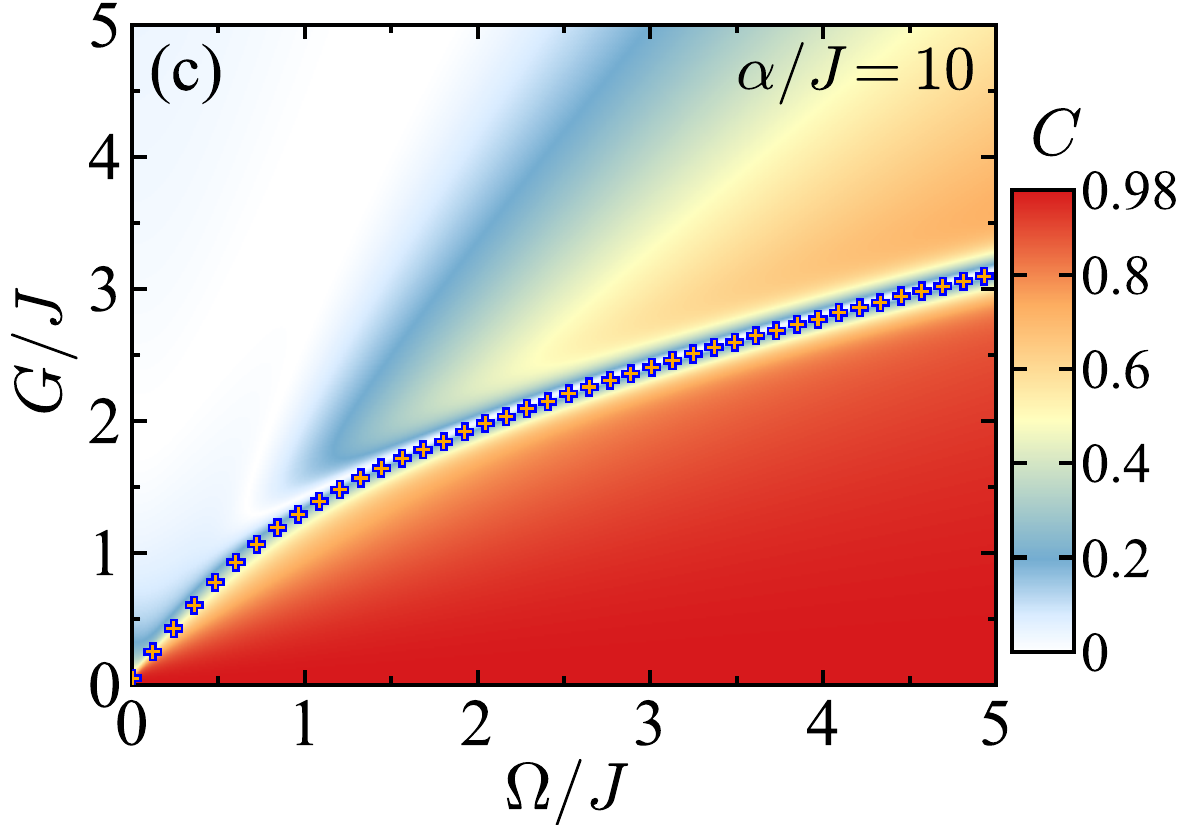}}
	\vspace{-0.25 cm}
	\caption{ Concurrence $C$ of the eDQD system in the $G/J$-$\Omega/J$ plane at $k_\text{B}T/J=0.01$ for (a) $\alpha/J=1$, (b) $\alpha/J=2$, and (c) $\alpha/J=10$. 
	The symbolic blue lines indicate the crossover boundary extracted from the maximum variation of the concurrence, $\max_G|\partial C/\partial G|$ at fixed $\Omega$ satisfying Eq. (\ref{Eq:symbols_abc}). This boundary closely coincides with the independently determined ground-state crossover associated with the onset of appreciable cavity-photon occupation. 
	 In three panels we have assumed, $B/J=2$, $N=5$ and $g_\text{c}/J=0$. }
	\label{fig:Con_GOmega_alpha}
\end{figure*}

Figure~\ref{fig:Con_GOmega_alpha} presents the concurrence in the $G/J$-$\Omega/J$ plane at the low temperature $k_{\rm B}T/J=0.01$ for three different  strengths of the Rashba SOC $\alpha/J=\{1, 2, 10\}$. We observe that the concurrence strongly depends on both the spin-photon coupling $G/J$ and the cavity frequency $\Omega/J$. The cavity does not simply produce a uniform renormalization of the concurrence, but it has the potential of generating distinct regions of enhanced and suppressed spin-charge entanglement in the $G/J-\Omega/J$ parameter space. 
For the case $\alpha/J=1$,  as shown in Fig.~\ref{fig:Con_GOmega_alpha}(a), the concurrence reaches values close to $C\simeq0.59$ in the strongly entangled region. The color map reveals that the concurrence changes substantially upon varying either $G/J$ or $\Omega/J$. In particular, a region in which the concurrence is strong appears on below the narrow crossover region, whereas the opposite side contains considerably weaker entanglement. This behavior demonstrates that the spin-photon interaction can reorganize the charge-spin composition of the low-energy dressed states and thereby strongly modify the entanglement such that it remains in the eDQD-dominated regime after the cavity is traced out. 

Increasing the Rashba coupling to $\alpha/J=2$, as shown in Fig.~\ref{fig:Con_GOmega_alpha}(b), on the one hand, leads to a substantial increase in the maximum concurrence with $C_\text{max}\simeq0.78$ at low temperature, and on the other hand, the region with strong entanglement becomes more prominent in the $G/J$-$\Omega/J$ plane. This result reflects the cooperative roles of the intrinsic spin-charge mixing generated by the Rashba term and the additional spin-photon hybridization induced by the cavity.  
The competition between Rashba SOC and Jaynes--Cummings term therefore allows the spin-charge correlations of the reduced eDQD state to be controlled by the electromagnetic environment. 

The effect of cavity QED on the maximum value of the spin--charge entanglement becomes less pronounced in the strong Rashba SOC $\alpha/J=10$ (see Fig.~\ref{fig:Con_GOmega_alpha}(c)). In this regime, the concurrence is already close to unity with a maximum value $C_\text{max}\simeq0.98$, consistent with the strong spin--charge entanglement reported for the cavity-free eDQD in Ref.~\cite{Ferreira2023}. Consequently, within the explored range of $G/J$ and $\Omega/J$, the cavity does not provide any further appreciable enhancement of the maximum entanglement, also it does not help us to drive the charge and spin degrees of freedom to a maximally entangled state with $C=1$. This behavior indicates that the cavity-induced enhancement becomes limited when the intrinsic Rashba interaction is sufficiently strong and the spin--charge entanglement is already close to saturation. Therefore, the cavity frequency and spin--photon coupling are most effective as external control parameters for tuning and enhancing the pairwise spin--charge entanglement at weak-to-moderate Rashba SOC regime, where they complement the intrinsic spin--charge hybridization generated by the Rashba interaction.

The nonmonotonic structure of the entanglement illustrated in Fig.~\ref{fig:Con_GOmega_alpha} is also important. As a matter of fact, increasing the light-matter coupling does not necessarily increase the reduced spin-charge entanglement monotonically. The cavity introduces an additional quantum degree of freedom, and part of the correlations can be redistributed among the charge, spin, and photon sectors. Depending on the values $G/J$, $\Omega/J$, and $\alpha/J$, this redistribution can either enhance or suppress the pairwise entanglement in the eDQD. 

A remarkable feature of Fig.~\ref{fig:Con_GOmega_alpha} is the well-defined nonlinear boundary shown by symbolic line. To clarify the physical origin of this boundary, we independently analyzed the ground state of the full eDQD--cavity Hamiltonian in the $(\Omega,G)$ plane. In particular, we monitored the ground-state photon occupation $n_{\rm ph}=\langle a^\dagger a\rangle$. The crossover from a predominantly eDQD-like state to a photon-dressed state occurs along a curve that closely follows the symbols superimposed on Fig.~\ref{fig:Con_GOmega_alpha}. These symbols were obtained independently from the position of the maximum variation of concurrence, $\max_G |\partial C/\partial G|$, at fixed $\Omega$. The coincidence of these two independently determined boundaries demonstrates that the significant variation of the concurrence is directly associated with a reorganization of the ground-state spin-charge-photon composition. Below the boundary, the photon occupation remains very small and the lowest-energy state is dominated by the eDQD degrees of freedom. Upon increasing the spin--photon coupling beyond a characteristic value $G_\text{c}(\Omega)$, the cavity acquires an appreciable occupation and the ground state becomes increasingly photon-dressed. 
Numerically, the crossover line is accurately described over the considered parameter interval by 
\begin{equation}\label{Eq:symbols_abc}
\frac{G_\text{c}}{J} = a\sqrt{\frac{\Omega}{J}} +b\frac{\Omega}{J}+c . 
\end{equation}
For $\alpha/J=1$, $2$, and $10$, respectively, we obtain $(a,b,c)=(1.4034, 0.0056,-0.0197)$, $(1.4593,-0.0079,-0.0877)$, and $(1.7956,-0.1008,-0.3927)$. The dominant square-root dependence, $G_\text{c}\propto\sqrt{\Omega}$, indicates that the crossover is governed by the competition between the photon excitation energy and the interaction-induced energy lowering associated with spin-photon hybridization. 

{ It is also noteworthy that the position of the crossover boundary changes only slightly with increasing Rashba SOC, whereas the concurrence is strongly enhanced. This behavior can be attributed to the stronger spin-charge hybridization induced by the Rashba interaction, which increases the nonseparable character of the charge and spin degrees of freedom and hence enhances their entanglement. At sufficiently strong SOC, the concurrence approaches unity, indicating an almost maximally entangled spin-charge state. The comparatively weak displacement of the crossover boundary suggests, however, that Rashba SOC primarily controls the magnitude of the spin-charge entanglement, while the location of the crossover remains governed predominantly by the competition between the cavity frequency and the spin-photon coupling.

The contour plots of the concurrence displaied in Fig. \ref{fig:Con_GOmega_alpha} therefore reflect the cooperative action of two distinct mechanisms: intrinsic spin-charge hybridization induced by the Rashba SOC (with low-to-moderate values) and cavity-induced entanglement controlled  by $G/J$ and $\Omega/J$. To verify that the location of the crossover lines are not artifact of the finite cavity Hilbert-space truncation, we repeated the ground-state calculation for several photon cutoffs such as $N = \{2, 4, 6, 8, 10, 20\}$. No appreciable displacement of the crossover boundary was observed within the resolution of the numerical grid . This convergence check confirms that the boundary identified from the photon occupation and the sharp variation of the concurrence is robust with respect to the cavity truncation in the parameter regime considered here.

\begin{figure}[t]
	\centering
	\resizebox{0.47\textwidth}{!}{
		\includegraphics[trim = 0 0 0 0, clip]{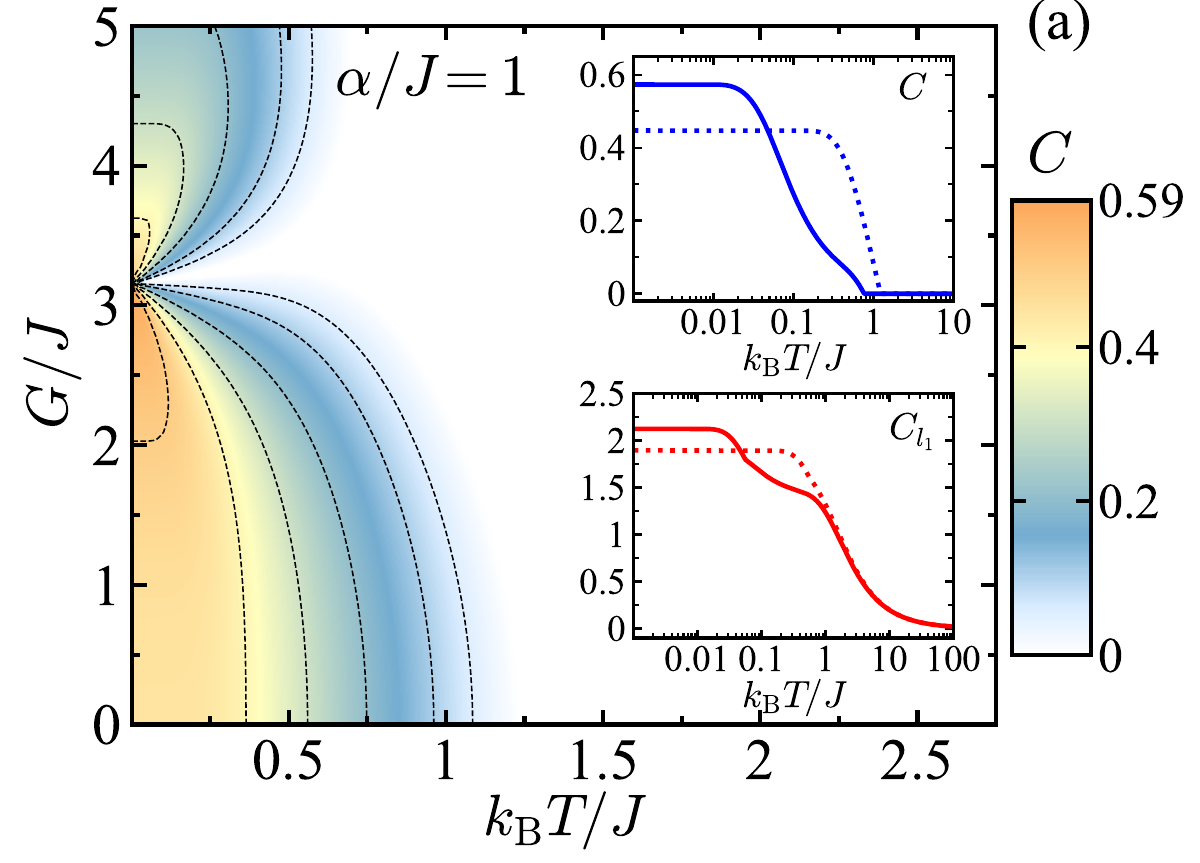}}
	\resizebox{0.47\textwidth}{!}{
		\includegraphics[trim = 0 0 0 0, clip]{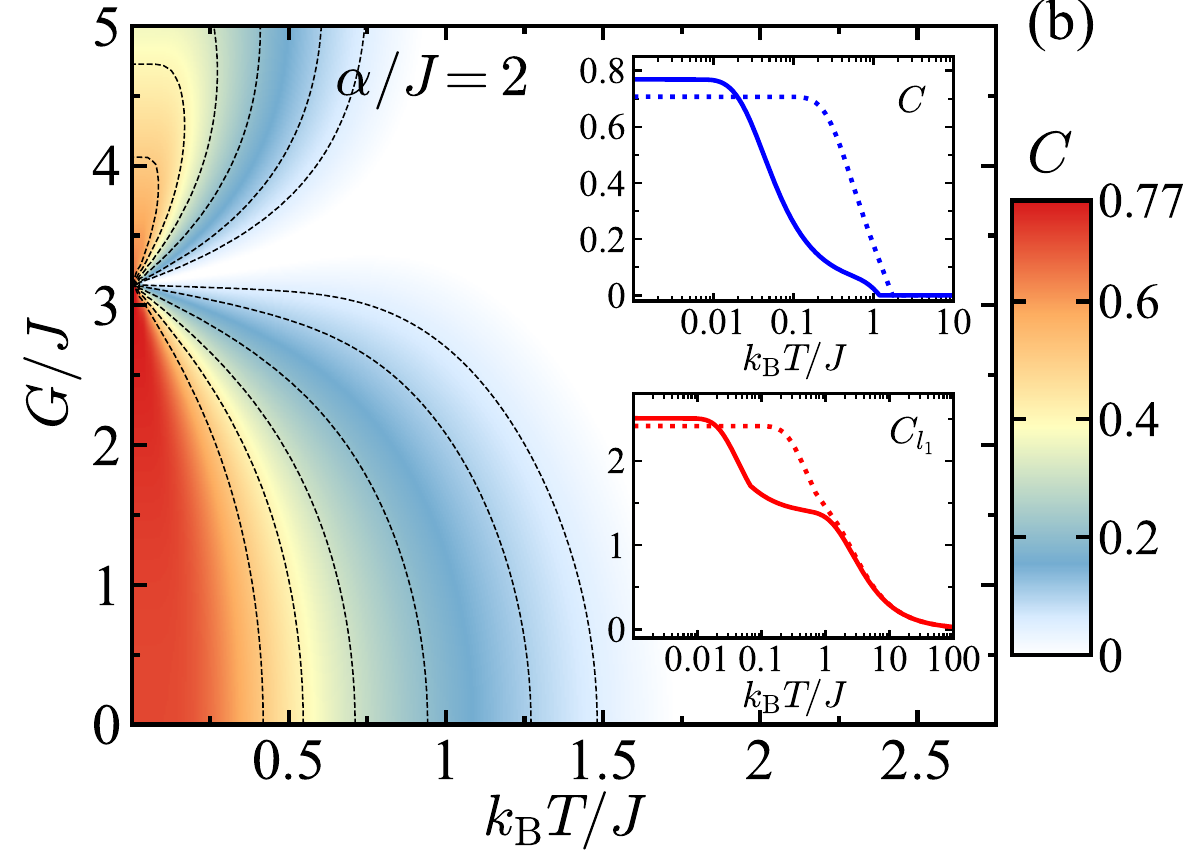}}
	\vspace{-0.25 cm}
	\caption{The concurrence of the eDQD system in the $G/J - T/J$ plane at $\Omega/J = 5$.
		(a) $\alpha / J = 1$, 
		(b) $\alpha / J = 2$.
		The upper insets represent the concurrence, while the lower insets show the $l_1$-norm coherence $C_{l_{1}}$, providing a complementary measure of the quantum coherence underlying the reduced eDQD state. The dotted curves in the insets show the corresponding results for the eDQD without the cavity with $G/J = \Omega/J = 0$, while the solid curves represent the cavity-coupled eDQD system with $G/J = 3$ and $\Omega/J = 5$. 
		In all panels, we have assumed: $B/J = 2$, $g_\text{c}/J = 0$, $N=5$.}
	\label{fig:ConCoh_GT_alpha}
\end{figure}

The role of thermal fluctuations emerged from temperature variations and the robustness of the quantum correlations in eDQD system are illustrated in Fig.~\ref{fig:ConCoh_GT_alpha}. The main panels show the concurrence in the $G/J-T/J$ plane at the fixed cavity frequency $\Omega/J=5$ for $\alpha/J=1$ and $2$. In both cases, the strongest concurrence is found in the low-temperature regime, where only the lowest photon-dressed states make a significant contribution to the thermal density matrix. Upon increasing the temperature, thermally excited states become populated and the reduced eDQD state becomes increasingly mixed, leading to a decay of the spin-charge entanglement. 

As shown in Fig.~\ref{fig:ConCoh_GT_alpha}(a), for $\alpha/J=1$, the concurrence exhibits a strong dependence on the spin-photon coupling $G/J$. The contour plot of the concurrence is divided into regions with substantially different sizes of the entanglement, separated by a narrow weakly entangled region around a characteristic coupling. This behavior confirms that the cavity can strongly restructure the reduced eDQD state even when the direct charge-photon coupling is absent ($g_{\rm c}/J=0$). The effect originates from the spin-photon Jaynes--Cummings interaction combined with the Rashba SOC. In this way, the cavity acts indirectly on the charge sector through the spin-charge hybridization. The corresponding result for $\alpha/J=2$ is displayed in Fig.~\ref{fig:ConCoh_GT_alpha}(b). The overall temperature dependence of the concurrence in the $G/J$-$T/J$ plane is retained, but the entanglement is enhanced at low temperature and its maximum value increases. The comparison between panels \ref{fig:ConCoh_GT_alpha}(a) and  \ref{fig:ConCoh_GT_alpha}(b) therefore demonstrates that despite the fact that stronger Rashba SOC favors the generation of spin-charge entanglement,  the cavity coupling provides an additional means of controlling where this entanglement is maximized. 

A particularly useful comparison with the cavity-free eDQD is provided by the upper insets of Fig.~\ref{fig:ConCoh_GT_alpha}. The dotted curves represent the corresponding results for the eDQD without the cavity, following the results reported in Ref.~\cite{Ferreira2023}, whereas the solid curves show the cavity-coupled eDQD system. For $\alpha/J=1$ (Fig. \ref{fig:ConCoh_GT_alpha}(a)), the concurrence at low temperature is increased from approximately $C\simeq0.45$ in the cavity-free case to about $C\simeq0.58$ in the presence of the cavity assuming $G/J = 3$ and $\Omega/J = 5$. For $\alpha/J=2$, a similar enhancement is observed, from approximately $C\simeq0.70$ to $C\simeq0.77$ with the same couplings to panel \ref{fig:ConCoh_GT_alpha}(b). Thus, cavity dressing produces a clear enhancement of the spin-charge entanglement in the low-temperature regime. Importantly, the comparison also shows that the cavity-induced enhancement is temperature dependent. Although the solid curves start from larger values of concurrence at low temperature, they begin to decrease as thermal excitations become relevant. The dotted and solid curves may therefore cross at intermediate temperatures, and the cavity-coupled concurrence does not remain larger throughout the complete temperature range. Hence, the principal cavity advantage revealed by the insets is an enhancement of the magnitude of the low-temperature entanglement rather than a universal increase of its thermal survival temperature.

The lower insets of Fig.~\ref{fig:ConCoh_GT_alpha} provide a complementary characterization in terms of the $l_1$-norm coherence $C_{l_1}$. For both values of the Rashba SOC $\alpha/J$, the cavity clearly increases the low-temperature coherence. For $\alpha/J=1$, $C_{l_1}$ is enhanced from approximately $1.9$ in the cavity-free system to about $2.1$ for $G/J=3$ and $\Omega/J=5$. Similarly, for $\alpha/J=2$, the low-temperature coherence increases from approximately $2.35$ to $2.5$. This enhancement is consistent with the increase in concurrence observed in the upper insets and shows that the cavity dressing modifies the eDQD state by strengthening its underlying quantum superposition. However, the behavior of $C_{l_1}$ also demonstrates that enhanced coherence should not be identified directly with enhanced entanglement, since coherence can include contributions that are not necessarily nonlocal. Moreover, the cavity-induced growth is most pronounced at low temperature: the solid red curves decrease at lower temperatures than the dotted curves and can cross them at intermediate temperatures. Hence, as for concurrence, the cavity primarily increases the magnitude of the quantum coherence in the low-temperature regime rather than universally increasing its thermal robustness. The different temperature dependencies emphasize that $C_{l_1}$ and $C$ are not equivalent measures. For instance, the cavity can generate significant coherence without that coherence necessarily translating into a corresponding increase in concurrence, particularly at very high temperatures.

\begin{figure}[b]
	\centering
	\resizebox{0.238\textwidth}{!}{
		\includegraphics[trim = 0 0 0 0, clip]{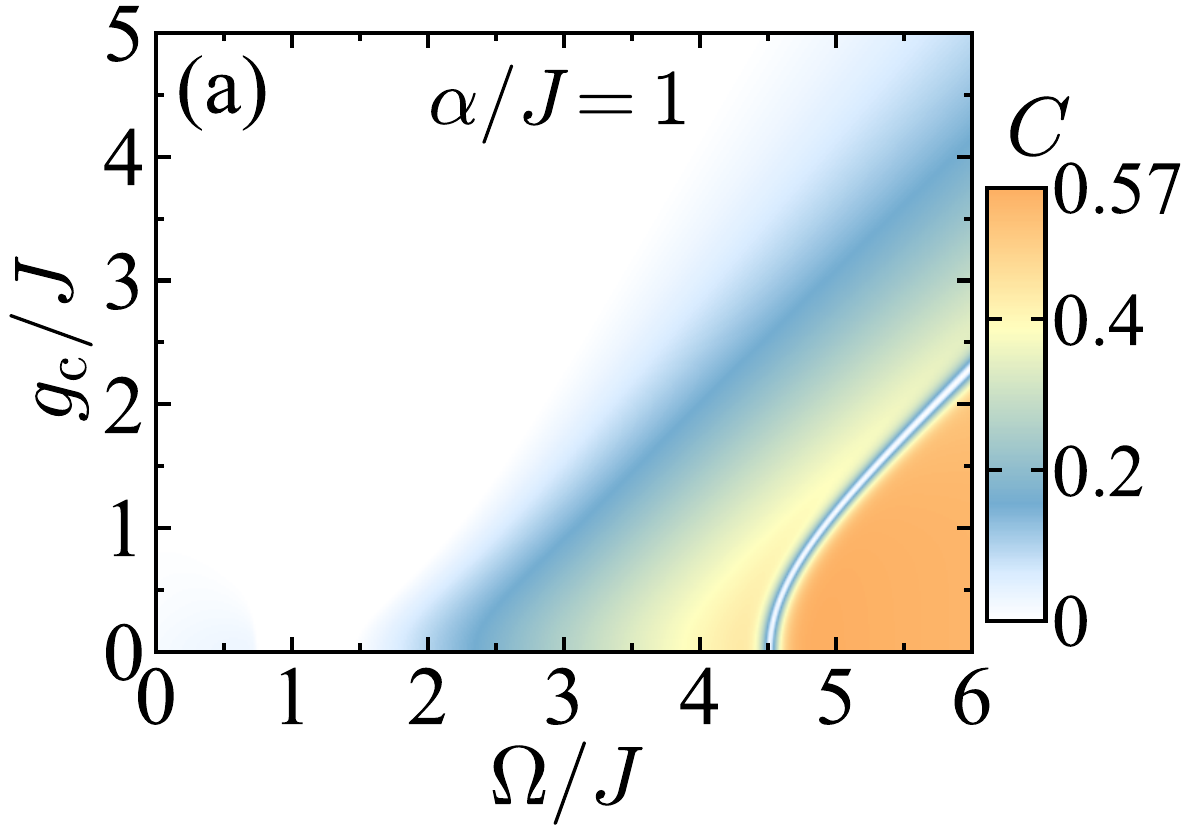}}
	\resizebox{0.238\textwidth}{!}{
		\includegraphics[trim = 0 0 0 0, clip]{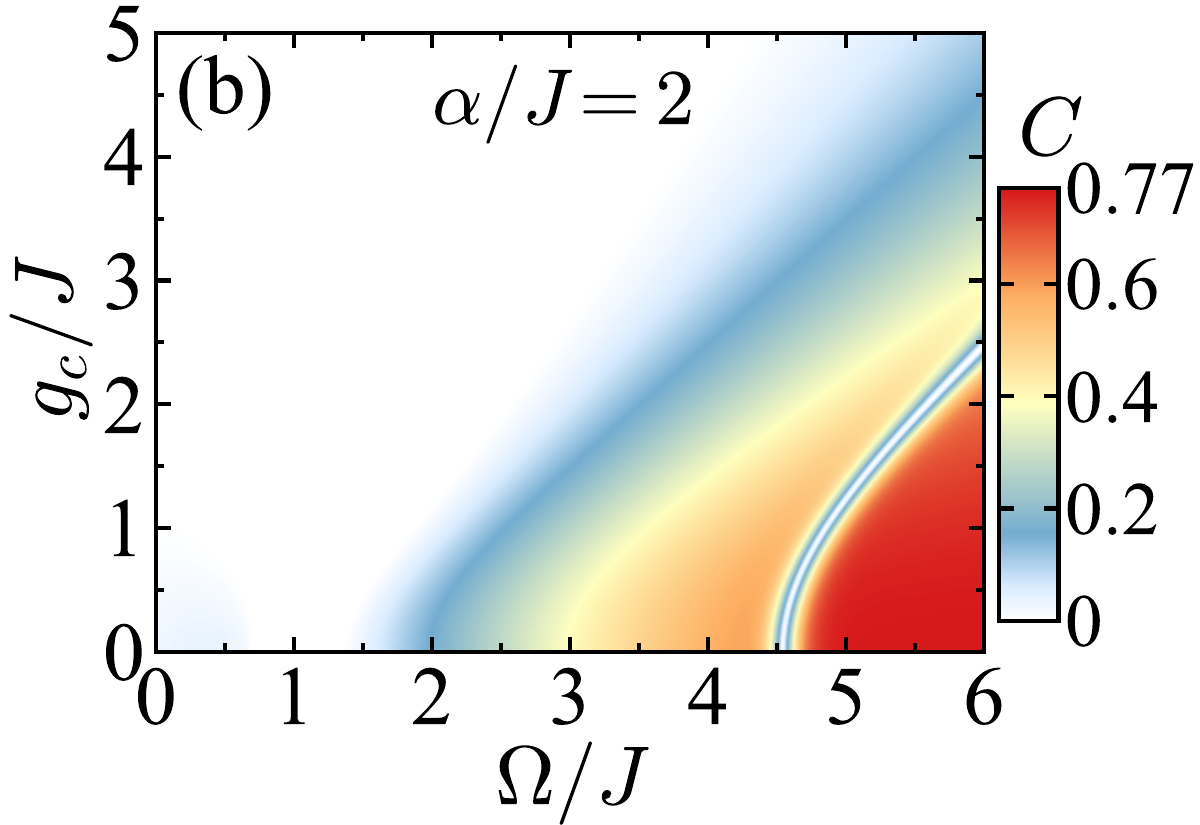}}
	\vspace{-0.7 cm}
	\caption{The concurrence of the eDQD system in the $  g_\text{c}/J-\Omega/J$ plane at $k_\text{B}T/J = 0.01$ and $G/J = 3$.
		(a) $\alpha / J = 1$, 
		(b) $\alpha / J = 2$.
		In both panels we have assumed: $B/J = 2$ and $N=5$.}
	\label{fig:Con_Ggc_alpha}
\end{figure}

The direct charge-photon coupling $g_\text{c}/J$ may play an important role in determining the degree of spin--charge entanglement in an experimentally realized cavity-coupled eDQD. It is therefore important to examine whether the enhancement of concurrence induced by the cavity persists in the presence of a finite $g_\text{c}/J$. To this end, we investigate the concurrence in the $g_\text{c}/J-\Omega/J$ plane at $k_{\rm B}T/J=0.01$ and $N=5$, fixing the spin-photon coupling at $G/J=3$, where a large concurrence is obtained for $g_\text{c}/J=0$ (e.g., see Fig.~\ref{fig:Con_GOmega_alpha}). For both $\alpha/J=1$ (Fig. \ref{fig:Con_Ggc_alpha})(a)) and $2$ (Fig. \ref{fig:Con_Ggc_alpha}(b)), our results show that the direct charge-photon coupling has a detrimental effect on the spin-charge entanglement. As $g_\text{c}/J$ increases, the concurrence progressively decreases and eventually vanishes, indicating that a strong direct coupling of the charge degree of freedom to the cavity is unfavorable for preserving the spin-charge entanglement. This suppression can be attributed to the increasing dressing of the charge degree of freedom by the cavity field, which redistributes the quantum correlations among the charge, spin, and photonic degrees of freedom. Consequently, after tracing out the cavity, the remaining pairwise entanglement between the spin and charge subsystems is reduced. Thus, while the spin-photon coupling $G/J$ can enhance the spin-charge concurrence in an appropriate parameter regime, a finite $g_\text{c}/J$ tends to counteract this enhancement, suggesting that weak direct charge-photon coupling is favorable for maintaining strong spin-charge entanglement in the eDQD.

\section{Cavity-mediated eDQD Battery}
\label{Sec:DQDs_QB}

The eDQD system provides a natural two-level QB in which the stored energy is associated with the internal charge and spin degrees of freedom of the electron. As illustrated in Fig.~\ref{fig:quantum_battery}, the proposed architecture combines an eDQD, a microwave cavity, and an auxiliary two-level charger.

Here, the cavity plays the role of a quantum mediator that transfers energy from the charger to the eDQD while also enabling coherent interactions between the charge and spin sectors of the battery. In this way, the charging process is not simply a direct classical energy injection into the DQD, but a cavity-mediated quantum energy-transfer process that can exploit quantum coherence and correlations to enhance energy storage and extraction.

The complete system shown in Fig \ref{fig:quantum_battery}, including the eDQD battery, the cavity mediator, and the charger, is governed by a modified Hamiltonian
\begin{equation}
	\hat{H}_{\text{M}}=\hat{H}_{\text{eDQD}}+\hat{H}_{\text{c}}+\hat{H}_{\text{eDQD-c}}+\hat{H}_{\text{ch-c}},
	\label{eq:total_QB_H}
\end{equation}
where the first three terms has the clear physical interpretation defined in Eq. \ref{Eq:total_H}.
 The fourth term describes the coupling of the cavity electric field to the charge dipole moment of the eDQD, which can induce charge excitations and provide an additional energy-storage channel.
On the other hand, the charger interacts with the cavity through a Jaynes--Cummings-type interaction,
\begin{equation}
\hat{H}_{\rm ch-c} = g_{\rm ch} \left( \hat{a}\hat{\sigma}^{\rm ch}_{+} + \hat{a}^{\dagger}\hat{\sigma}^{\rm ch}_{-} \right),
\label{eq:charger_cavity}
\end{equation}
where $\hat{\sigma}^\text{ch}_{\pm}$ are the raising and lowering operators of the charger, and $g_{\rm ch}$ is the charger--cavity coupling strength. An initially excited charger can therefore transfer its energy coherently to the cavity, creating microwave photons that subsequently interact with the eDQD.

\begin{figure}[t]
		\centering
	\resizebox{0.49\textwidth}{!}{
		\includegraphics[trim = 10 0 0 0,
	clip]{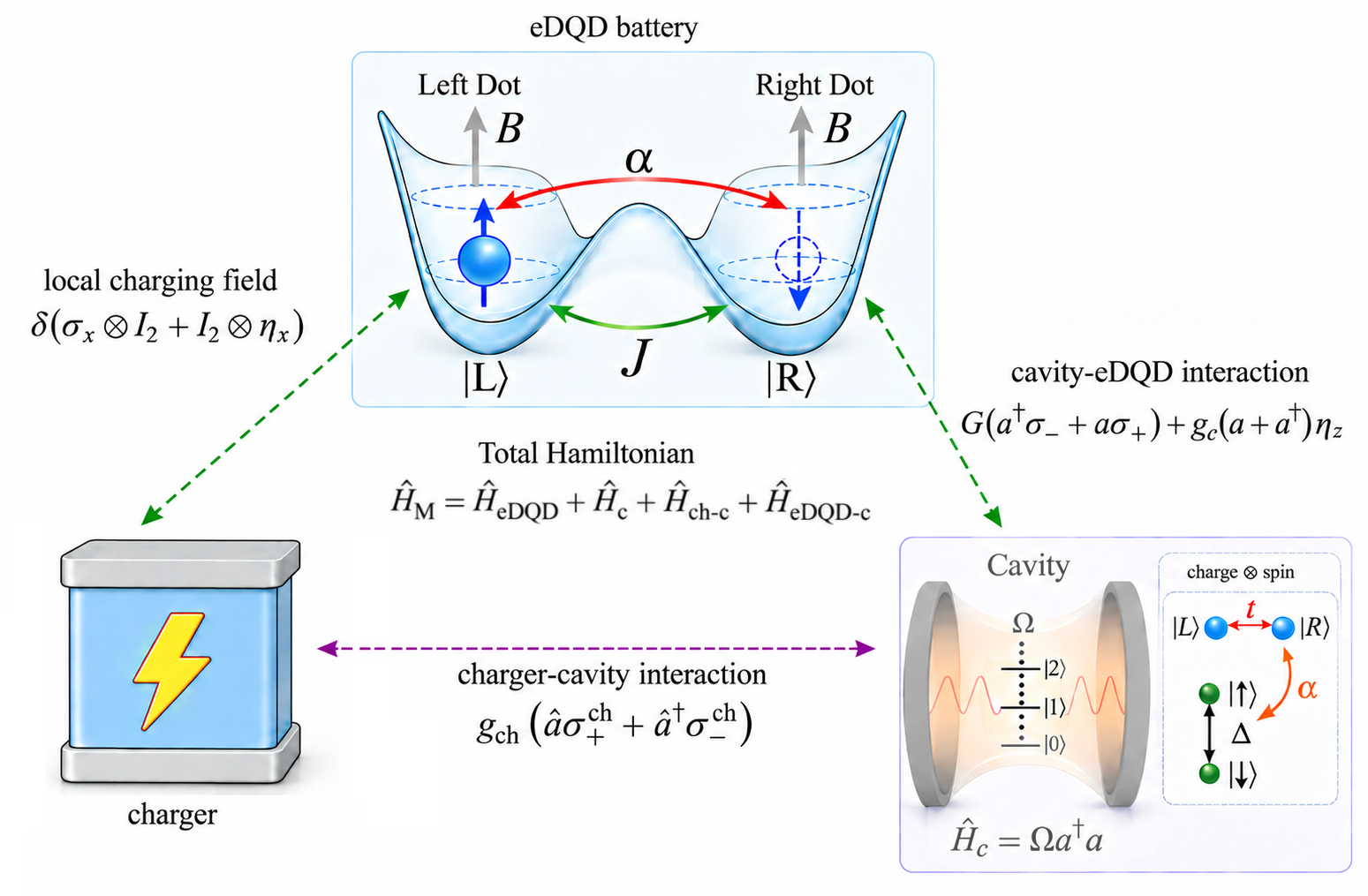}}
	\caption{
		Schematic representation of the cavity-mediated eDQD QB. The eDQD, comprising coupled charge and spin degrees of freedom, constitutes the QB, while a single-mode microwave cavity mediates the interaction between the battery and the charging process. The charge and spin sectors are coupled through the Rashba SOC with strength $\alpha$, while $J$ denotes the inter-dot tunneling amplitude and $B$ the detuning energy. The cavity couples to the DQD spin degree of freedom through the Jaynes--Cummings interaction with strength $G$, and to the charge degree of freedom through $g_\text{c}$. The battery can be charged via the cavity-mediated process or by a local external magnetic field applied along either the $x$- or $y$-direction, with charging strength $B$.
	}
	\label{fig:quantum_battery}
\end{figure}

\subsection{The charging protocol}
\label{subsec:charging_protocol}

The eDQD QB can be charged either directly by a local external field that couples to the charge or spin degrees of freedom, or indirectly through the cavity-mediated interaction with the auxiliary charger. In this work, we consider both routes and focus on the unitary charging dynamics generated by a local magnetic field applied uniformly to the eDQD.
A general local charging field acting on the eDQD can be written as
\begin{equation}
	\hat{H}_{\rm ch} = \omega_\text{ch}(t)\left[{\sigma}_{(x,y)} \otimes \mathbb{I} + \mathbb{I} \otimes {\eta}_{(x,y)}\right],
	\label{eq:local_charging_general}
\end{equation}
where \(\omega_\text{ch}(t)\) is the charging field strength, which may be time-dependent or time-independent. For simplicity, we assume a time-independent charging field, \(\omega_\text{ch}(t)=\omega_\text{ch}\), unless stated otherwise. The Pauli operators \({\sigma}_{(x,y)}\) and \({\eta}_{(x,y)}\) act on the spin and charge subspaces, respectively. Thus, the field couples simultaneously to the spin and charge degrees of freedom of the eDQD, driving transitions between the corresponding energy levels of the battery.

Both Pauli-\(X\) and Pauli-\(Y\) gates can equivalently serve as charging operators for the QB. The Pauli-\(X\) gate acts as a pure bit-flip operation, while the Pauli-\(Y\) gate performs a combined bit-flip and phase-flip operation. Since both are unitary, they can drive the battery from its initial state to a charged state. The Pauli-\(X\) charging is particularly suitable when only population transfer is required, whereas the Pauli-\(Y\) charging provides additional phase control and can be advantageous when the battery configuration or the field interactions require phase-sensitive manipulation. In the presence of Rashba SOC and cavity-mediated interactions, this additional phase degree of freedom can significantly influence the extractable work and the charging efficiency.
For a closed system, the charging process is implemented as a unitary evolution
\begin{equation}
	\hat{U}_{\rm ch}(t) = \exp\left(-i\hat{H}_{\rm ch}t\right),
	\label{eq:unitary_charging}
\end{equation}
which governs the cyclic charging of the QB over time. At the maximum stored energy, reached at time \(t=\tau\), the charging field must be disconnected to prevent the battery from reverting to an uncharged state due to continuous cycling. In fact, the charging protocol is completed by disconnecting the external field at the optimal charging time $t=\tau$, defined as the time at which the ergotropy attainable under the chosen charging dynamics reaches its maximum. Since the evolution generated by $\hat{H}_{\rm ch}$ is restricted to a specific unitary trajectory, the state reached at $t=\tau$ does not necessarily coincide with the active state associated with the initial density-matrix spectrum. Only when ${\rho}(\tau)={\rho}_{\rm act}$ does the battery attain the maximum ergotropy allowed by its unitary orbit. The charging field is therefore switched off at $t=\tau$ to preserve the maximum extractable work attained along the chosen charging trajectory and to prevent its subsequent decrease due to the cyclic unitary evolution.

During the charging phase, we apply the unitary operator in Eq.~\eqref{eq:unitary_charging}, $\hat{U}_{\rm ch}(t) = \exp(-i\hat{H}_{\rm ch} t)$, which evolves the QB from its initial passive Gibbs state $\hat{\zeta}$ to the time-dependent state
\begin{equation}
	\hat{\rho}(t)
	=\hat{U}_{\rm ch}(t)\,{\rho_\text{eDQD}}(T)\,
	\hat{U}^{\dagger}_{\rm ch}(t),
	\label{eq:rho_t_under_charging}
\end{equation}
where $\rho_\text{eDQD}(T)$ is defined in Eq. (\ref{eq:Reduced_rho_eDQD}). Since $\rho_\text{eDQD}(T)$ is diagonal in the eigenbasis of $\hat{H}_{\rm eDQD}$ with populations that decrease monotonically with energy, it is a passive state from which no work can be extracted by any cyclic unitary operation. At certain evolution times, the charged state $\hat{\rho}(t)$ may coincide with the active state
\begin{equation}
\rho_\text{eDQD}^\text{act}= \sum_{\mu} \phi_{\mu}\, \lvert \psi_{\mu} \rangle \langle \psi_{\mu} \rvert,
\qquad
\phi_{\mu+1} \geq \phi_{\mu} \;\;\text{for all}\;\; \mu , 
\label{eq:active_state_eta}
\end{equation}
which is obtained by assigning the largest eigenvalues $\phi_{\mu}$ of $\hat{\rho}(t)$ to the highest-energy eigenstates $\lvert \psi_{\mu} \rangle$ of the battery Hamiltonian $\hat{H}_{\rm eDQD}$ (\ref{eq:H_eDQD}). In such cases, the stored energy corresponds to the maximum extractable work discussed in the next part. This alignment is not generic. Namely, the charged state reaches the active state is determined by the specific form of the charging unitary $\hat{U}_{\rm ch}(t)$ and by the spectrum of the battery Hamiltonian $\hat{H}_{\rm eDQD}$ at the chosen operating point.

To make the charging protocol explicit, we express the charging Hamiltonian in the four-dimensional basis of the eDQD charge-spin subspace defined in Eq. (\ref{Eq:eDQD_basis}).
For a charging field applied along the \(y\)-direction, the charging Hamiltonian in the ordered basis (\ref{Eq:eDQD_basis}) takes the form
\begin{equation}
	\hat{H}^{(y)}_{\rm ch}
	=
	\omega_\text{ch}
	\begin{pmatrix}
		0 & -i & -i & 0 \\
		i & 0 & 0 & -i \\
		i & 0 & 0 & -i \\
		0 & i & i & 0
	\end{pmatrix}.
	\label{eq:Hy_matrix}
\end{equation}
The corresponding unitary evolution operator is therefore
\begin{equation}
	\hat{U}^{(y)}_{\rm ch}(t)
	=
	\exp\left(-i\hat{H}^{(y)}_{\rm ch}t\right)
	=
	\begin{pmatrix}
		\mathcal{A} & \mathcal{P} & \mathcal{P} & \mathcal{Q} \\
		\mathcal{P} & \mathcal{A} & \mathcal{Q} & \mathcal{P} \\
		\mathcal{P} & \mathcal{Q} & \mathcal{A} & \mathcal{P} \\
		\mathcal{Q} & \mathcal{P} & \mathcal{P} & \mathcal{A}
	\end{pmatrix},
	\label{eq:Uy_matrix}
\end{equation}
where $\mathcal{A} = \cos^2(\omega_\text{ch} t),  \mathcal{Q} = -\sin^2(\omega_\text{ch} t), 
	\mathcal{P} = -\frac{i}{2}\sin(2\omega_\text{ch} t)$.
This unitary operator clearly shows the cyclic nature of the charging process: the diagonal elements \(\mathcal{A}\) describe the return amplitude of the initial populations, while the off-diagonal elements \(\mathcal{Q}\) and \(\mathcal{P}\) describe coherent population transfer and phase accumulation among the charge-spin states. In particular, the presence of both off-diagonal elements reflects the combined bit-flip and phase-flip character of the Pauli-\(Y\) charging field.

For a charging field applied along the \(x\)-direction, the charging Hamiltonian is instead
\begin{equation}
	\hat{H}^{(x)}_{\rm ch}=	\omega_\text{ch}
	\left(\hat{\sigma}_x \otimes \mathbb{I}
	+\mathbb{I} \otimes \hat{\eta}_x\right),
	\label{eq:Hx_charging}
\end{equation}
where the unitary operator \(\hat{U}^{(x)}_{\rm ch}\) has the same structural form as Eq.~\eqref{eq:Uy_matrix}, but with real coefficients~\cite{HaddadiAQT2025b}.
Although the matrix forms of \(\hat{U}^{(x)}_{\rm ch}\) and \(\hat{U}^{(y)}_{\rm ch}\) are identical in the basis (\ref{Eq:eDQD_basis}), the physical interpretation differs: the Pauli-\(X\) charging generates purely real bit-flip amplitudes, whereas the Pauli-\(Y\) charging generates imaginary amplitudes associated with phase flips. This distinction becomes important when the charging field is combined with the Rashba SOC and the cavity-mediated interactions, since these terms are sensitive to the relative phases between the charge and spin components.

The charging protocol is completed by disconnecting the external field at the optimal charging time \(t=\tau\), defined as the time at which the stored energy of the battery reaches its maximum. At this instant, the unitary evolution is stopped and the battery is decoupled from the charging source. The maximum stored energy and the corresponding optimal charging time depend on the interplay among the tunneling amplitude \(J\), the Zeeman splitting \(B\), the Rashba SOC \(\alpha\), the cavity coupling strengths \(G\) and \(g_{\rm ch}\), and the charging strength \(\omega_\text{ch}\). These dependencies are analyzed numerically in the following sections.

It is also important to distinguish the direct local charging protocol described here from the cavity-mediated charging protocol introduced in Eq.~\eqref{eq:charger_cavity}. In the latter, the charger transfers energy to the cavity, which then interacts with the eDQD through both spin and charge channels. The local charging field, by contrast, acts directly on the eDQD and therefore provides a useful benchmark for quantifying the enhancement or suppression of energy storage due to the cavity. In both cases, the charging process is unitary and coherent, so that the initial state, the charging Hamiltonian, and the coherence properties of the eDQD battery ultimately limit the maximum extractable work.

\subsection{Performance Indicators for the eDQD Battery}

The reduced density matrix defined in Eq.~\eqref{eq:Reduced_rho_eDQD} contains all the information needed to compute the stored energy and ergotropy of the eDQD battery. To quantify the performance of the eDQD system as a QB, we consider two key measures: the stored energy and the ergotropy.

The stored energy is defined relative to the initial battery state as
\begin{equation}
	E_{\rm st}(t) = \mathrm{Tr}\!\left[ \rho_{\rm eDQD}(t)\hat{H}_{\rm eDQD} \right] - \mathrm{Tr}\!\left[ \rho_{\rm eDQD}(0)\hat{H}_{\rm eDQD} \right],
	\label{eq:stored_energy}
\end{equation}
where \(\rho_{\rm eDQD}(t)\) is the reduced density matrix of the eDQD system at time \(t\). This quantity measures the total energy accumulated by the eDQD system during the charging process. However, not all of this energy is necessarily extractable as useful work.

The ergotropy quantifies the maximum amount of useful work that can be extracted from the eDQD battery through a cyclic unitary process. For a battery state \(\rho_{\rm eDQD}\) and Hamiltonian \(\hat{H}_{\rm eDQD}\), the ergotropy is defined as
\begin{align}
	\mathcal{E}(\rho_{\rm eDQD}) &= \mathrm{Tr}\left(\rho_{\rm eDQD}\hat{H}_{\rm eDQD}\right) \nonumber\\
	& - \min_{U} \mathrm{Tr} \left( U\rho_{\rm eDQD}U^\dagger \hat{H}_{\rm eDQD} \right), \label{eq:ergotropy}
\end{align}
where the minimization is performed over all unitary transformations \(U\) acting on the battery. The second term represents the energy of the corresponding passive state, i.e., the minimum energy that can be reached through unitary operations. Equivalently, the ergotropy can be expressed as
\begin{equation}
	\mathcal{E} = \mathrm{Tr}\left[(\rho_{\rm eDQD} - \rho_{\rm eDQD}^{\rm pas})\hat{H}_{\rm eDQD}\right],
\end{equation}
where \(\rho_{\rm eDQD}^{\rm pas}\) is the passive state obtained by rearranging the populations of \(\rho_{\rm eDQD}\) in increasing order with the energy levels.

The passive state is uniquely determined by the battery Hamiltonian, up to degeneracies in its energy spectrum. For the present eDQD QB, the relevant Hamiltonian is the full isolated eDQD Hamiltonian \(\hat{H}_{\rm eDQD}\). Let its spectral decomposition be
\begin{equation}
	\hat{H}_{\rm eDQD} = \sum_{m=1}^{4} \varepsilon_m |\psi_m\rangle\langle\psi_m|, \qquad \varepsilon_1 \leq \varepsilon_2 \leq \varepsilon_3 \leq \varepsilon_4,
	\label{eq:DQD_spectral_decomposition}
\end{equation}
where \(|\psi_m\rangle\) are the energy eigenstates of the eDQD system. If the reduced density matrix of the battery is written in its spectral decomposition as
\begin{equation}
	\rho_{\rm eDQD} = \sum_{m=1}^{4} r_m |r_m\rangle\langle r_m|,
\end{equation}
with the eigenvalues ordered according to
\begin{equation}
	r_1 \geq r_2 \geq r_3 \geq r_4,
\end{equation}
then the passive state is obtained by assigning the largest population to the lowest-energy eigenstate, the next-largest population to the next-lowest energy eigenstate, and so forth. Thus,
\begin{equation}
	\rho_{\rm eDQD}^{\rm pas} = \sum_{m=1}^{4} r_m |\psi_m\rangle\langle\psi_m|.
	\label{eq:passive_state_DQD}
\end{equation}

Equivalently, the passive state is obtained when the eigenvectors of \(\rho_{\rm eDQD}\) are aligned with the energy eigenbasis of \(\hat{H}_{\rm eDQD}\) according to
\begin{equation}
	|r_m\rangle = |\psi_m\rangle, \qquad m = 1, \ldots, 4,
\end{equation}
with populations decreasing monotonically as the energy increases. Consequently, the passive state satisfies
\begin{equation}
	\langle\psi_m|\rho_{\rm eDQD}^{\rm pas}|\psi_m\rangle \geq \langle\psi_{m+1}|\rho_{\rm eDQD}^{\rm pas}|\psi_{m+1}\rangle,
\end{equation}
for \(\varepsilon_m < \varepsilon_{m+1}\).

The minimum energy accessible through cyclic unitary transformations is therefore
\begin{equation}
	E_{\rm pas} = \mathrm{Tr}\left[ \rho_{\rm eDQD}^{\rm pas} \hat{H}_{\rm eDQD} \right] = \sum_{m=1}^{4} r_m \varepsilon_m,
	\label{eq:passive_energy_eDQD}
\end{equation}
and the ergotropy can be written explicitly as
\begin{equation}
	\mathcal{E}(\rho_{\rm eDQD}) = \mathrm{Tr}\left[ \rho_{\rm eDQD}\hat{H}_{\rm eDQD} \right] - \sum_{m=1}^{4} r_m \varepsilon_m.
	\label{eq:ergotropy_passive_DQD}
\end{equation}

It is important to note that, in the present eDQD system, the passive state is generally not identical to the computational-basis ground state or to a state defined simply by the localized charge and spin configurations.
\begin{figure*}[t]
	\centering
	\resizebox{0.327\textwidth}{!}{
		\includegraphics[trim = 0 0 0 0, clip]{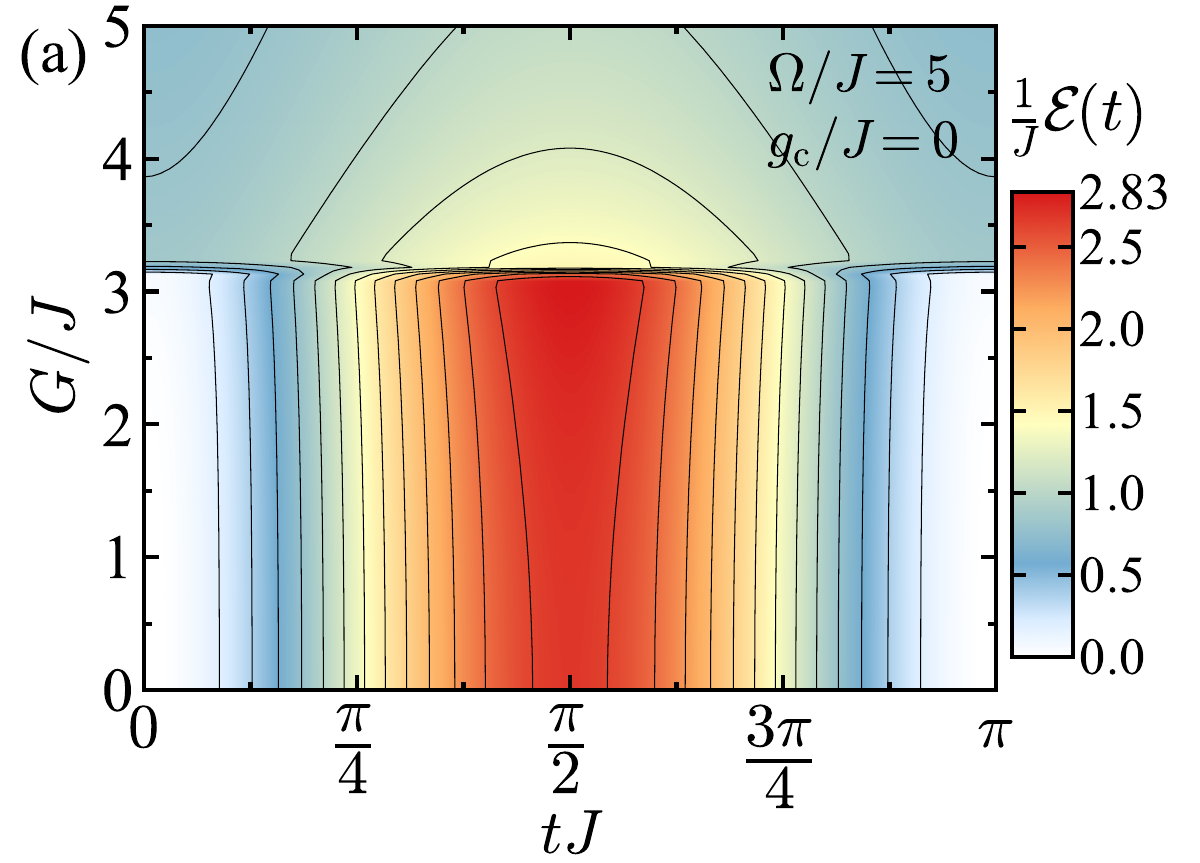}}
	\resizebox{0.327\textwidth}{!}{
		\includegraphics[trim = 0 0 0 0, clip]{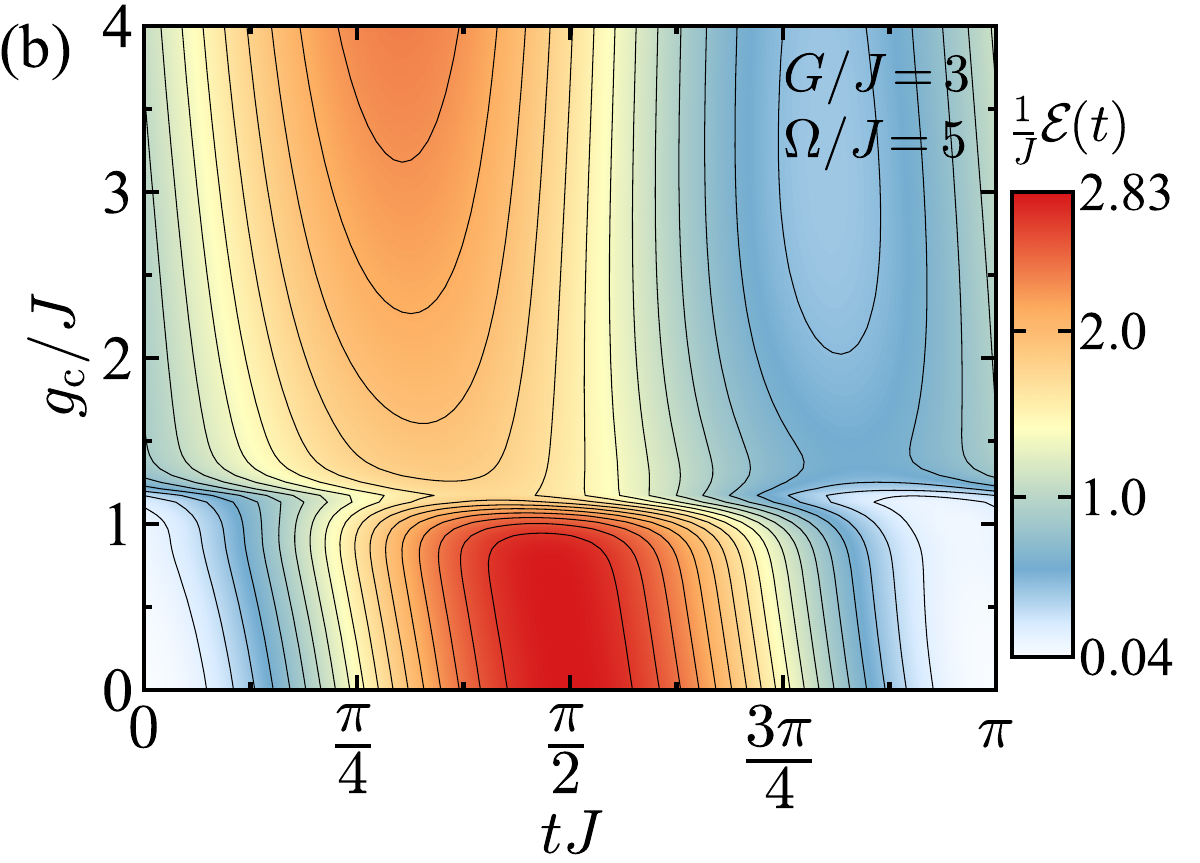}}
	\resizebox{0.327\textwidth}{!}{
		\includegraphics[trim = 0 0 0 0, clip]{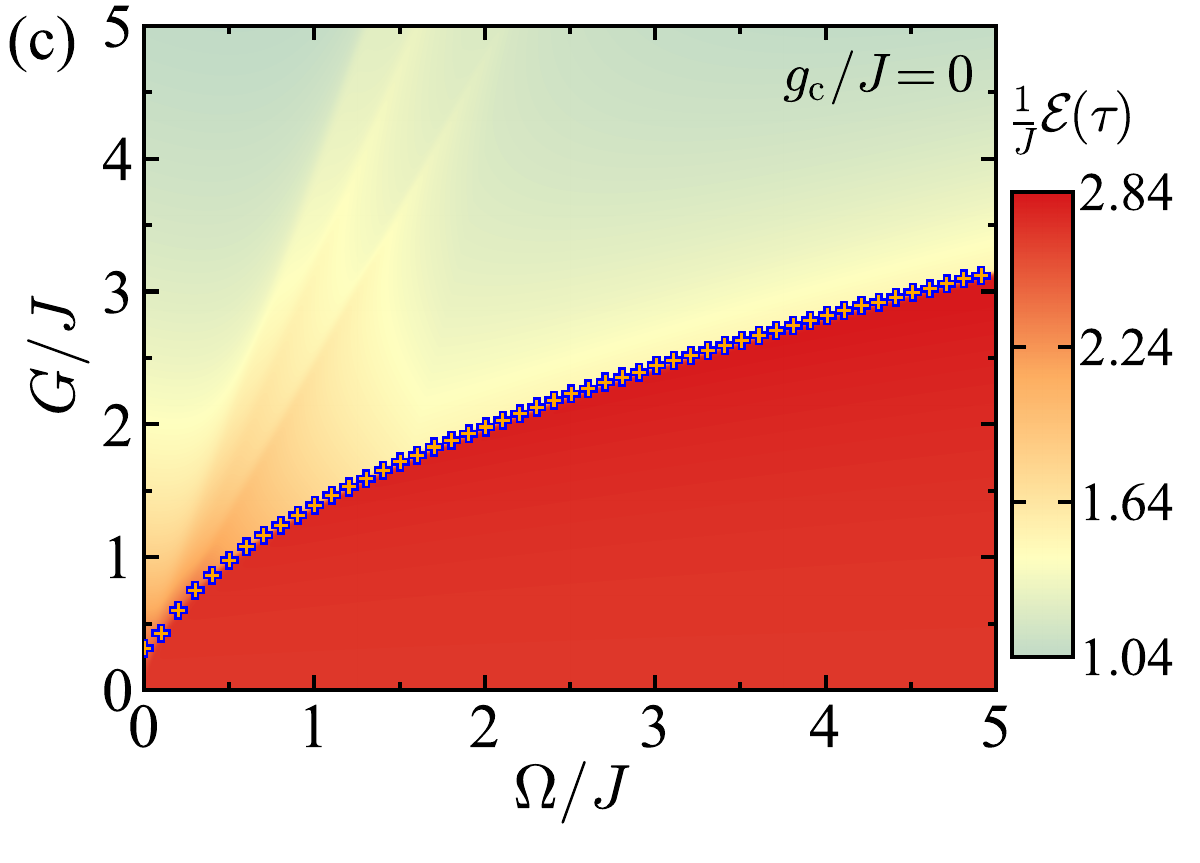}}
    \resizebox{0.327\textwidth}{!}{
		\includegraphics[trim = 0 0 0 0, clip]{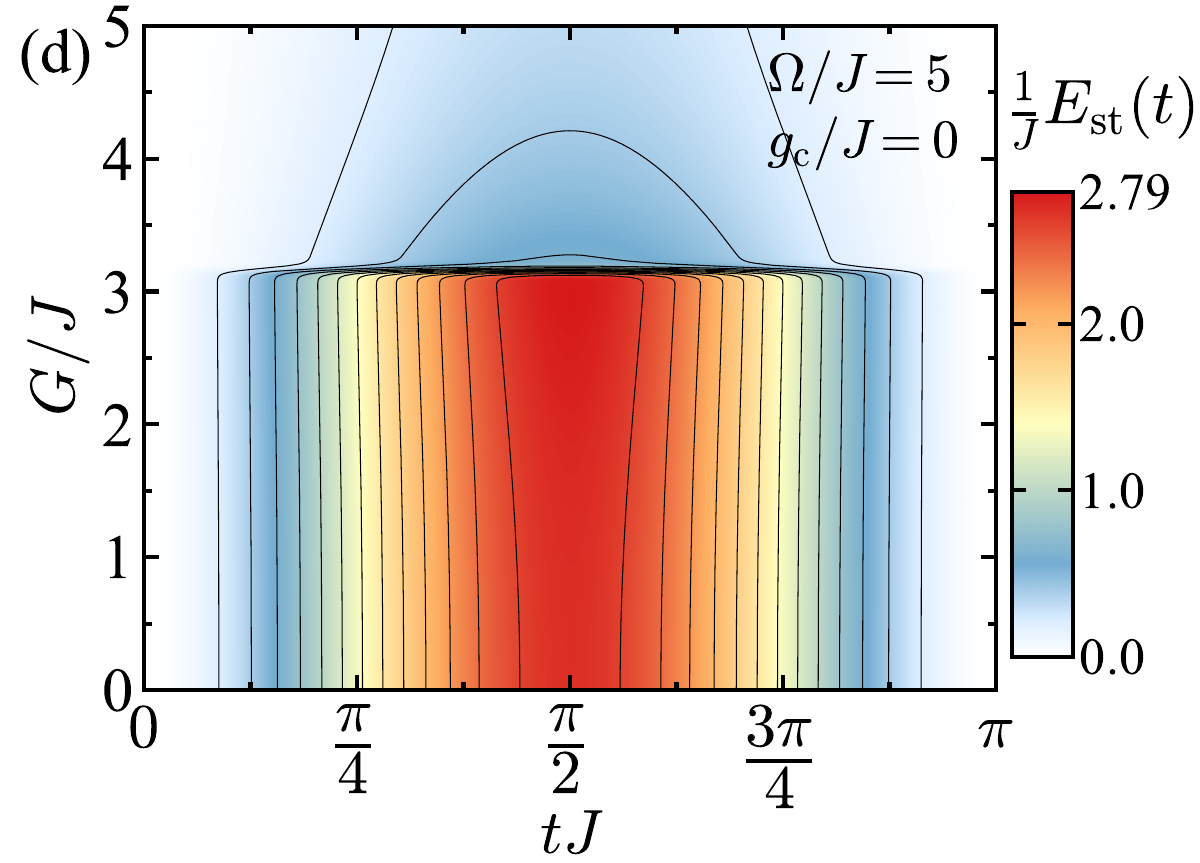}}
	\resizebox{0.327\textwidth}{!}{
		\includegraphics[trim = 0 0 0 0, clip]{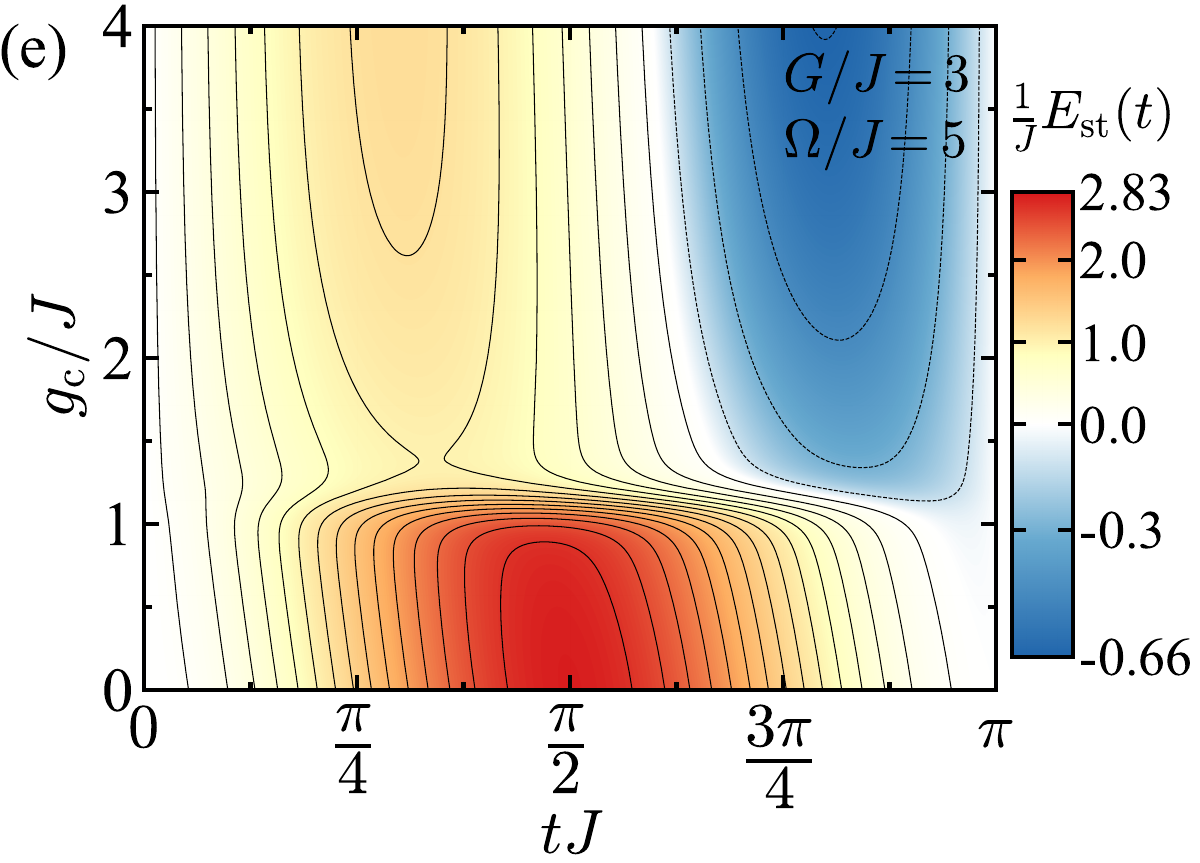}}
	\resizebox{0.327\textwidth}{!}{
		\includegraphics[trim = 0 0 0 0, clip]{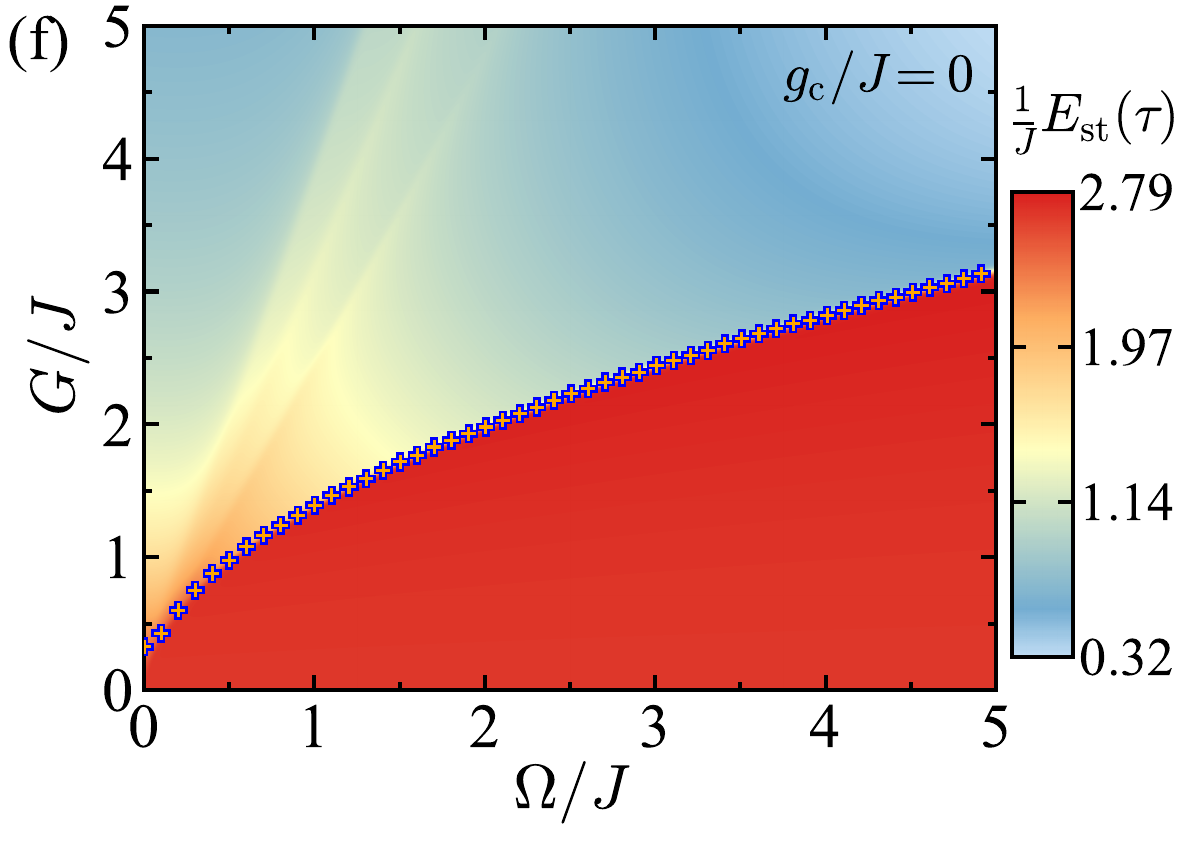}}
	\vspace{-0.25 cm}
	\caption{(a) and (d): The ergotropy and stored energy in the $G/J - tJ$ plane for fixed values $\Omega/J=5$ and $g_\text{c}/J = 0$. (b) and (e): The ergotropy and stored energy in the $g_\text{c}/J-tJ$ plane for $G/J=3$ and $\Omega/J=5$. (c) and (f): Ergotropy and stored energy in the $G/J-\Omega/J$ plane for $g_\text{c}/J = 0$.  In all three panels we supposed $\alpha/J=1$, $k_\text{B}T/J = 0.01$, $B/J=2$, and $N = 5$.}
	\label{fig:Ergo_gcGOmega_alpha1}
\end{figure*}
Because the tunneling term \(J\eta_x\) and the Rashba SOC \(\alpha\eta_y\sigma_x\) hybridize the charge and spin degrees of freedom, the eigenstates \(|\psi_m\rangle\) of \(\hat{H}_{\rm eDQD}\) are generally superpositions of the localized spin-charge basis states. The passive state must therefore be constructed from the actual energy spectrum and eigenstates of \(\hat{H}_{\rm eDQD}\) at the parameters used in the charging protocol.

For the thermal density matrix in Eq.~(\ref{eq:Reduced_rho_eDQD}), the populations decrease monotonically with increasing energy. Hence, the thermal state is passive and satisfies \(\rho_{\rm eDQD}(T) = \rho_{\rm eDQD}^{\rm pas}\). This provides the connection between the thermodynamic reference state and the unitary definition of ergotropy used for the eDQD QB.

\subsection{Results}
Figure~\ref{fig:Ergo_gcGOmega_alpha1} illustrates how the extractable work of the eDQD battery is controlled by the cavity parameters and by the charging time. 
The charging dynamics are resolved explicitly in Figs.~\ref{fig:Ergo_gcGOmega_alpha1}(a) and \ref{fig:Ergo_gcGOmega_alpha1}(b). For $g_{\rm c}/J=0$ and $\Omega/J=5$, as shown in panel~\ref{fig:Ergo_gcGOmega_alpha1}(a), a notable oscillatory dependence of the ergotropy on the dimensionless charging time $tJ$ is observed that in turn reflects the coherent and cyclic character of the unitary charging protocol. The ergotropy is strongly enhanced within selected temporal windows, with a broad maximum centered approximately around $tJ\simeq\pi/2$, whereas considerably smaller values occur near the beginning and end of the charging cycle. The dependence on $G/J$ changes qualitatively around the same coupling region identified from the static $G/J$--$\Omega/J$ map. Consequently, the spin-photon interaction controls not only the magnitude of the available extractable work but also the temporal profile through which this work is generated during coherent charging. On the other hand, panel~\ref{fig:Ergo_gcGOmega_alpha1}(b) depicts the additional influence of the direct charge-photon coupling $g_{\rm c}/J$ at fixed $G/J=3$ and $\Omega/J=5$ (close to the boundary crossover). We see different behavior of the ergotropy when $g_\text{c}/J\neq 0$ compared to the simple temporal oscillation obtained for $g_{\rm c}/J=0$. In fact, a finite charge-photon interaction substantially restructures the ergotropy pattern. The position and magnitude of the high-ergotropy region become strongly dependent on $g_{\rm c}/J$, indicating that the charge-photon channel provides an additional control parameter for the battery dynamics. Particularly large ergotropy is obtained within restricted regions of the $(g_{\rm c}/J,tJ)$ plane rather than through a monotonic increase with $g_{\rm c}/J$. This nonmonotonic behavior emphasizes that stronger light--matter coupling does not automatically imply improved battery performance, instead, the useful extractable energy results from the combined action of the spin-photon coupling, direct charge-photon coupling, Rashba-induced spin-charge hybridization, and the phase accumulated during the coherent charging evolution.

Panel~\ref{fig:Ergo_gcGOmega_alpha1}(c) shows the ergotropy in the $G/J$--$\Omega/J$ plane for $g_{\rm c}/J=0$. A particular feature is the nonlinear boundary separating two regimes with outstandingly different ergotropy. Below this boundary, the ergotropy remains comparatively large and is weakly dependent on the spin-photon coupling $G/J$, whereas crossing the boundary toward larger $G/J$ produces an abrupt reduction and a stronger dependence on the cavity parameters. In particular, the location and shape of this boundary closely resemble the previously identified crossover from the concurrence (\ref{Eq:symbols_abc}) and from the ground-state photon occupation in Fig.~\ref{fig:Con_GOmega_alpha}. The crossover extracted from the ergotropy for $\alpha/J=1$ is well represented by
\begin{equation}
	\frac{G_{\rm c}}{J}=1.4295\sqrt{\frac{\Omega}{J}}-0.0028\frac{\Omega}{J}-0.0338 .
\end{equation}
The close correspondence between this crossover and the sharp variation of the ergotropy demonstrates that the cavity-induced reorganization of the low-energy eDQD--photon states is manifested not only in the spin-charge entanglement but also in the amount of useful work that can be extracted from the reduced eDQD battery. In the eDQD-dominated regime below the crossover, the cavity occupation remains small and the reduced battery retains a structure that is comparatively favorable for the considered charging protocol. Beyond the crossover, stronger photon dressing modifies the populations and coherences of the reduced eDQD state after the cavity is traced out, leading to a substantial redistribution of its extractable energy. Thus, almost similar cavity-dressing crossover that governs the sharp variation of concurrence also provides a characteristic boundary for the energetic performance of the eDQD battery. This correspondence should not, however, be interpreted as a direct proportionality between concurrence and ergotropy: concurrence quantifies pairwise spin-charge entanglement, whereas ergotropy measures the maximum work extractable by unitary operations. Rather, both quantities provide complementary signatures of the underlying restructuring of the cavity-dressed eDQD state.

Panels~\ref{fig:Ergo_gcGOmega_alpha1}(d)-\ref{fig:Ergo_gcGOmega_alpha1}(f) show the stored energy \(\tfrac{1}{J}E_{\rm st}(t)\) that quantifies the total energy deposited in the eDQD--cavity system during charging, in the same three parameter planes. Comparing the stored energy with the ergotropy brings insights that the two quantities share the same qualitative structure, showing the same sharp crossover curve, and very similar periodic \(tJ\)-striping. This is the indication of a battery operating near the coherent-charging optimum, where almost all the injected energy is extractable, so \(\mathcal{E} \approx E_{\rm st}\) and the ergotropy tracks the stored energy closely. The numerical difference between the two is the passive energy \(E_{\rm pas} = E_{\rm st} - \mathcal{E}\), which is the part of the stored energy that cannot be extracted by any unitary operation because it resides in the completely passive (thermal) state. More strikingly, in the red regions of Figs.~\ref{fig:Ergo_gcGOmega_alpha1}(a)--\ref{fig:Ergo_gcGOmega_alpha1}(c), this passive contribution is negligible, whereas in the pale and blue regions it clearly dominates.

The most significant feature of the stored energy is appeared in Fig.~\ref{fig:Ergo_gcGOmega_alpha1}(e), where it has a negative value, ranging down to \(-0.66\) (dark blue), in contrast to Fig.~\ref{fig:Ergo_gcGOmega_alpha1}(b) where ergotropy tends to zero. A negative stored energy means the eDQD--cavity system has less energy than its reference thermal state at the same temperature. As a matter of fact, the charging protocol has driven the system below its equilibrium energy. This is physically possible because the stored energy is defined relative to the initial thermal state of the uncoupled eDQD. When the charge-photon coupling \(g_{\rm c}\) is strong enough to modify the energy spectrum, the cavity-mediated spin-charge-photon interaction can cool the system. The dressed states then lie below the bare thermal mixture in energy, which makes the stored energy negative. The deep blue trough of negative stored energy in Fig.~\ref{fig:Ergo_gcGOmega_alpha1}(e), located at \(tJ \gtrsim 3\pi/4\) and \(g_{\rm c}/J > 1.0\), coincides with the region where the ergotropy in Fig.~\ref{fig:Ergo_gcGOmega_alpha1}(b) has already dropped to below unity. Under these circumstances, the system has been driven into a state that is more passive than the bare thermal mixture, so no work can be extracted and energy must instead be supplied to return the eDQD--cavity system to equilibrium.

The connection between stored energy and ergotropy is particularly important. First, in the regimes where the battery charges effectively (\(G/J\lesssim 3\), small \(g_{\rm c}/J\)), stored energy and ergotropy are nearly equal, and both are maximized at \(tJ = \tau J = \pi/2\), confirming that coherent charging deposits energy in an almost fully extractable form. Second, in the strong light-matter coupling regimes where stored energy goes negative, ergotropy is also suppressed, because the cavity dressing has pushed the system into a passive-like state with no useful work content. The crossover curve in Figs.~\ref{fig:Ergo_gcGOmega_alpha1}(c) and~\ref{fig:Ergo_gcGOmega_alpha1}(f) is the same in both quantities---\(G \propto \sqrt{\Omega}\)---demonstrating that the restructuring of the dressed spin-charge-photon states controls both how much energy is stored and how much of it can be extracted. This tight correlation between \(\mathcal{E}\) and \(E_{\rm st}\) across all three parameter planes establishes the eDQD--cavity system as a coherent quantum battery whose charging performance is dictated by the same cavity-dressing physics that governs its quantum correlations.

We also find that raising the Rashba coupling to \(\alpha/J = 2\) boosts the stored energy considerably, so that stronger spin-orbit hybridization improves not only the extractable work but also the total energy deposited during charging. To keep the presentation concise, we have shown detailed plots only for \(\alpha/J = 1\), and the corresponding results for \(\alpha > 1\) are straightforwardly achievable  by repeating the same calculations with the larger Rashba coupling, and are therefore omitted here.

To determine whether the direct charge-photon interaction can be exploited to optimize the battery performance, we next maximize the ergotropy with respect to $g_{\rm c}/J$ at every point of the $(G/J,tJ)$ plane. The results are displayed in Fig.~\ref{fig:MaxErgo_3D} for $\alpha/J=1$ and $2$. The optimization shown in the figure is performed over the interval $0\leq g_{\rm c}/J\leq4$. We have additionally verified the robustness of the optimized ergotropy by extending the upper bound of the search interval to $g_{\rm c}^{\max}/J=5$, $8$ and higher. These extensions produce no vivid change in the resulting maximum ergotropy profile  $\mathcal{E}_{\max}(G,t)$, demonstrating that the reported distribution of the optimized extractable work is almost insensitive to the choice $g_{\rm c}^{\max}/J=4$ over the investigated parameter range. In Fig.~\ref{fig:MaxErgo_3D}, the vertical coordinate represents the normalized optimal charge-photon coupling $g_{\rm c}^{\rm opt}/g_{\rm c}^{\max}$ selected by the maximization procedure for each pair $(G/J,tJ)$, whereas the surface color represents the corresponding maximum ergotropy $\mathcal{E}_{\max}/J$. The figure therefore simultaneously provides information about the charge-photon coupling favored by the optimization and the maximum extractable work attainable at each operating point.

For $\alpha/J=1$, as shown in Fig.~\ref{fig:MaxErgo_3D}(a), the coupling selected by the optimization exhibits a highly structured dependence on both the charging time $tJ$ and the spin-photon coupling $G/J$. Extended regions in which the optimum occurs at or near the upper boundary, $g_{\rm c}^{\rm opt}/g_{\rm c}^{\max}\simeq1$, are separated by sharp changes toward regions where substantially weaker charge-photon coupling is favored. These abrupt variations indicate switching between competing maxima of the ergotropy as $G/J$ and $tJ$ are varied. In regions where $g_{\rm c}^{\rm opt}/g_{\rm c}^{\max}=1$, the result should be interpreted as indicating that the largest coupling within the considered optimization interval is favored, rather than as establishing a finite global optimum at $g_{\rm c}/J=4$. Nevertheless, extending $g_{\rm c}^{\max}/J$ beyond $4$ leaves the corresponding maximum-ergotropy landscape essentially unchanged, confirming that $\mathcal{E}_{\max}$ itself is robust with respect to the numerical cutoff. Accordingly, a single fixed value of $g_{\rm c}/J$ cannot optimize the battery over the complete charging cycle. Moreover, the surface colors show that a large value of $g_{\rm c}^{\rm opt}/g_{\rm c}^{\max}$ does not necessarily correspond to a large $\mathcal{E}_{\max}$. The coupling favored by the optimization and the amount of extractable work obtained at that optimum therefore provide two distinct pieces of information about the battery performance.

\begin{figure}[t]
	\centering
	\resizebox{0.45\textwidth}{!}{
		\includegraphics[trim = 100 50 100 120, clip]{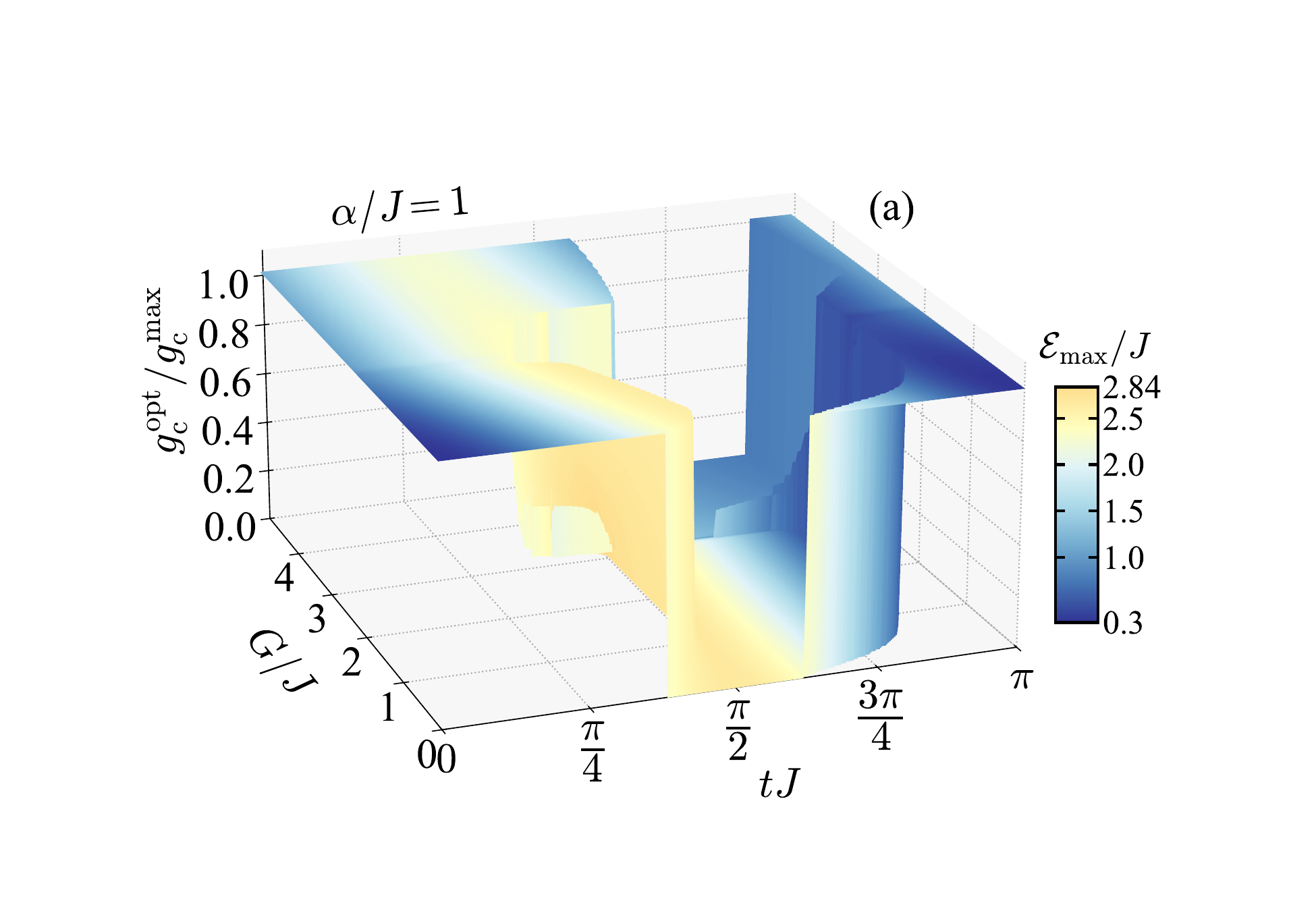}}
	\resizebox{0.45\textwidth}{!}{
		\includegraphics[trim = 100 50 100 120, clip]{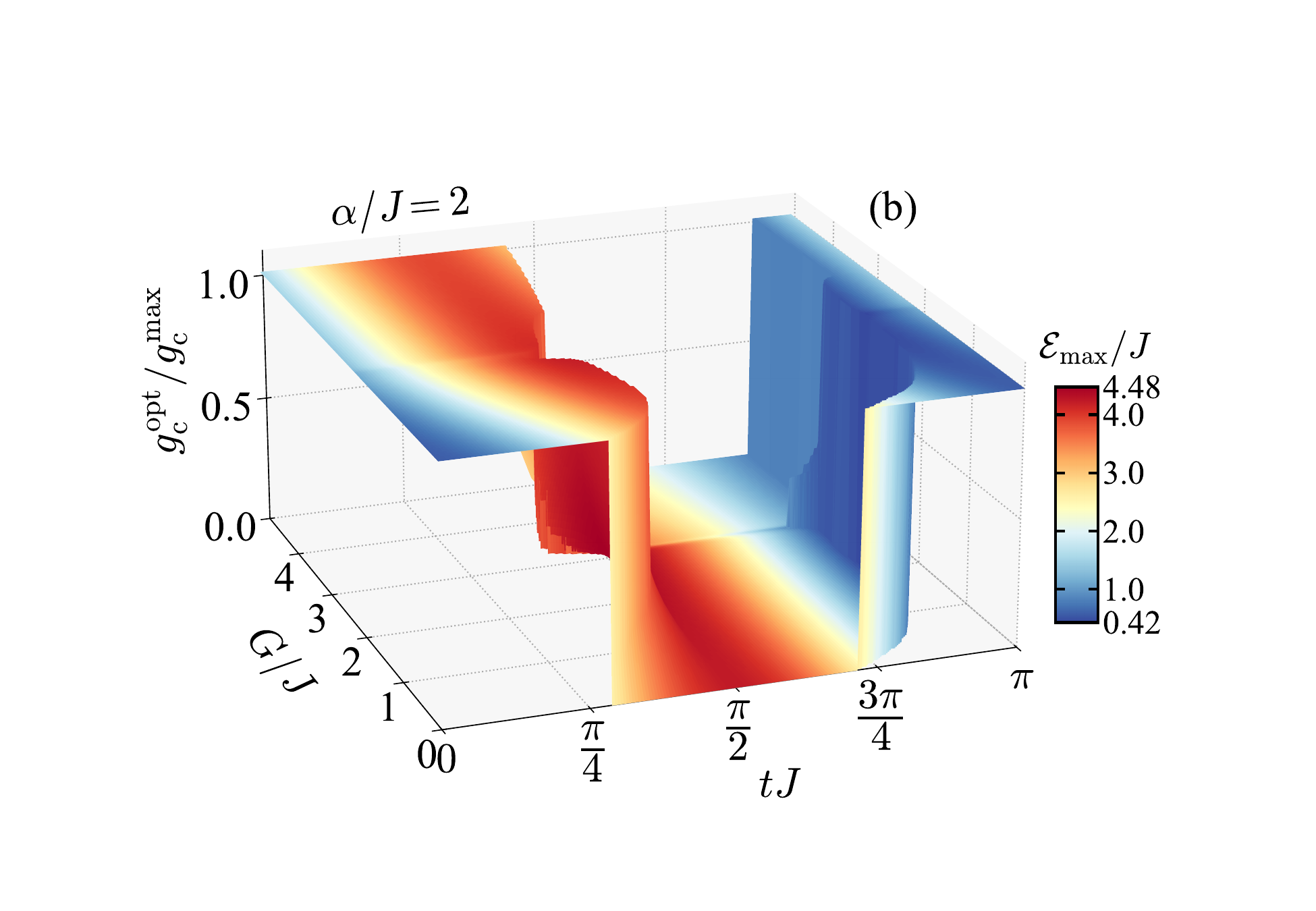}}
	\vspace{-0.25 cm}
	\caption{Maximum ergotropy $\mathcal{E}_{\max}/J$ achieved within the $(g_{\mathrm{c}}^{\mathrm{opt}}/g_{\mathrm{c}}^{\mathrm{max}}; G/J; tJ)$ grid for (a) $\alpha/J = 1$ and (b) $\alpha/J = 2$. We have assumed $k_\text{B}T/J = 0.01$, $B/J=2$, $\Omega/J = 5$ and $N = 5$. }
	\label{fig:MaxErgo_3D}
\end{figure}
As shown in Fig.~\ref{fig:MaxErgo_3D}(b), increasing the Rashba coupling to $\alpha/J=2$ preserves the overall configuration of the optimal coupling $g_{\rm c}^{\rm opt}/J$  but substantially increases the accessible ergotropy, with the maximum value rising from approximately $\mathcal{E}_{\max}/J\simeq2.84$ for $\alpha/J=1$ to  $\mathcal{E}_{\max}/J\simeq4.48$ for $\alpha/J=2$. This enhancement demonstrates that stronger intrinsic spin-charge hybridization can increase the amount of useful work made available by the charging operation. Moreover, the persistence of the sharp variations in $g_{\mathrm{c}}^{\mathrm{opt}}$ indicates that Rashba SOC does not remove the competition between the different cavity-dressed charging regimes. It instead renormalizes the energetic scale, leaving the optimal charge-photon coupling strongly dependent on $G/J$ and on the charging time $tJ$.
In result, Figs.~\ref{fig:Ergo_gcGOmega_alpha1} and \ref{fig:MaxErgo_3D} demonstrate that the cavity plays two complementary roles in the eDQD battery. First, the spin-photon interaction reshape the state of the reduced eDQD across similar boundary crossover that is visible in the concurrence, thereby modifying the extractable work. Second, the direct charge-photon interaction provides an additional optimization tool whose favorable value depends on the charging time and on the spin-photon coupling. The battery performance is therefore governed not simply by maximizing an individual coupling strength, but by selecting an appropriate combination of $G/J$, $g_{\rm c}/J$, $\Omega/J$, $\alpha/J$, and charging time.

It is instructive to compare these energetic findings with the quantum-resource analysis presented in Sec.~\ref{Sec:ConCoh}. The most direct correspondence occurs between Fig.~\ref{fig:Ergo_gcGOmega_alpha1}(a) and Fig.~\ref{fig:Con_GOmega_alpha}(a), where both quantities undergo a sharp alteration across similar nonlinear boundary in the $(G/J, \Omega/J)$ plane. In the concurrence analysis, this boundary was independently associated with the crossover from an eDQD-dominated low-energy state to an increasingly photon-dressed state through the onset of appreciable cavity occupation. The appearance of almost similar structure in the ergotropy shows that this cavity-induced reconfiguration has consequences beyond quantum correlations and directly affects the energetic usefulness of the reduced eDQD state. Nevertheless, concurrence, coherence, and ergotropy quantify different properties and do not need to be maximized simultaneously. Concurrence measures specifically the pairwise entanglement between the spin and charge degrees of freedom, the $l_1$-norm of coherence quantifies basis-dependent quantum coherence of the reduced eDQD state, whereas ergotropy measures its departure from passivity and hence the work extractable through cyclic unitary operations. The common crossover observed in these quantities should therefore be regarded as evidence that they respond to the same underlying change in the spin-charge-photon composition of the dressed states, rather than as evidence of a direct one-to-one relation between entanglement, coherence, and extractable work.

\section{Conclusion}\label{sec:conclusion}

In this paper, we have presented a comprehensive theoretical investigation of quantum coherence, spin-charge entanglement, and QB performance in a single-electron silicon DQD coupled to a microwave cavity. Starting from a microscopic Hamiltonian that incorporates interdot tunneling, Rashba SOC, Zeeman splitting, the cavity mode, Jaynes--Cummings spin-photon coupling, and direct charge-photon interaction, we constructed the thermal state of the composite eDQD--cavity system by tracing out the photonic degrees of freedom, and characterized the reduced eDQD state through its concurrence, the \(l_1\)-norm of coherence, and its charging properties. In the introduced model, the cavity acts as an effective control knob for the quantum resources of the eDQD subsystem. At low temperature, the spin-photon coupling and cavity frequency strongly modify the concurrence of the spin and charge, producing distinct regions of enhanced and suppressed bipartite entanglement in the \((\Omega/J, G/J)\) plane, while the coherence exhibits complementary behavior. A well-defined cavity-dressing crossover emerges whose boundary scales as \(G_{\mathrm{c}} \propto \sqrt{\Omega}\), coincides with the onset of appreciable ground-state photon occupation, and is robust against variations of the cavity Hilbert-space truncation. The direct charge-photon coupling \(g_{\mathrm{c}}\) predominantly suppresses the concurrence, revealing that the spin-photon and charge-photon channels play qualitatively different roles. 

Treating the eDQD as a QB under coherent local charging, the ergotropy and the stored energy both exhibit a considerable change across the same crossover, with the ergotropy closely tracking the stored energy in the efficiently charging regime and deviating from it once the passive contribution grows. Remarkably, when the charge-photon coupling \(g_{\mathrm{c}}\) is large, the stored energy can become negative. This means the coupled eDQD-cavity system holds less energy than the uncoupled eDQD at thermal equilibrium, so the cavity has effectively cooled the system. Optimizing the ergotropy with respect to \(g_{\mathrm{c}}/J\) yields a highly structured optimal-coupling profile with sharp transitions between charging regimes, in which a larger \(g_{\mathrm{c}}^{\mathrm{opt}}\) does not necessarily imply a larger maximum ergotropy. We also understood that increasing the Rashba SOC substantially enhances the accessible ergotropy. 

In summary, our results establish a direct link between cavity-controlled quantum resources and the energetic response of the single electron confined in a DQD. The cavity frequency, spin-photon coupling, charge-photon coupling, Rashba SOC, and charging time form a complementary set of control parameters that reshape the reduced eDQD state and thereby tune its coherence, spin-charge entanglement, extractable work, and stored energy. The appearance of qualitatively the same cavity-dressing crossover in both the entanglement and ergotropy surfaces provides direct evidence that cavity QED modifies both the quantum correlations and the work-extraction capability of the eDQD. These findings position cavity-embedded semiconductor DQDs as a promising theoretical platform for exploring the interplay between spin-charge quantum resources, light--matter hybridization, and quantum energy storage, and motivate future work for incorporating dissipative dynamics, cavity losses, charge and spin decoherence, and experimentally accessible time-dependent charging protocols.



\bibliographystyle{apsrev4-2}
\bibliography{eDQD_bibFile}

\end{document}